\documentclass[prd,twocolumn,superscriptaddress,floatfix,amsmath,amssymb,amsfonts,longbibliography,nofootinbib]{revtex4-2}

\usepackage{float} 

\usepackage[normalem]{ulem}
\usepackage[english]{babel}
\usepackage{graphicx}
\usepackage{dcolumn}
\usepackage{bm}
\usepackage{blindtext}
\usepackage{verbatim}
\usepackage{relsize}
\usepackage{mathrsfs}
\usepackage{musicography}
\usepackage{amsmath}
\usepackage{blindtext}
\usepackage{cancel}
\usepackage{physics}
\usepackage{epstopdf}
\usepackage{mathtools}
\usepackage{tensor}
\usepackage{color}
\usepackage{tikz}
\usepackage[usenames,dvipsnames]{pstricks}
\usepackage{epsfig}
\usepackage{pst-grad} 
\usepackage{pst-plot} 
\usepackage{hyperref}
\hypersetup{
    colorlinks=true,
    linkcolor=blue,
    filecolor=magenta,      
    urlcolor=cyan,
}
\usepackage{verbatim}
\usepackage{slashed}

\usepackage{dsfont}

\newcommand{\ts}{\textsc}

\newcommand{\rhoh}{\hat{\rho}}
\newcommand{\phih}{\hat{\phi}}
\newcommand{\proj}[2]{\ket{#1} \! \bra{#2}}
\newcommand{\muh}{\hat{\mu}}
\newcommand{\mf}{\mathsf}

\newcommand{\ii}{\mathrm{i}}

\newcommand{\M}{\mathcal{M}}
\renewcommand{\L}{\mathcal{L}}

\usepackage{xcolor}

\newcommand{\jpg}[1]{{\color{orange} \bf [Jos\'{e}: #1]}}

\allowdisplaybreaks[1] 

\begin{document}

\title{How measuring a quantum field affects entanglement harvesting}

\author{H\'ector Maeso-Garc\'{i}a}
\email{hmaesoga@uwaterloo.ca}
\affiliation{Centre de Formaci\'o Interdisciplin\`{a}ria Superior, Universitat Polit\`{e}cnica de Catalunya, 08028 Barcelona, Spain}
\affiliation{Department of Applied Mathematics, University of Waterloo, Waterloo, Ontario, N2L 3G1, Canada}

\author{Jos\'e Polo-G\'omez} 
\email{jpologomez@uwaterloo.ca}
\affiliation{Department of Applied Mathematics, University of Waterloo, Waterloo, Ontario, N2L 3G1, Canada}
\affiliation{Institute for Quantum Computing, University of Waterloo, Waterloo, Ontario, N2L 3G1, Canada}
\affiliation{Perimeter Institute for Theoretical Physics, Waterloo, Ontario, N2L 2Y5, Canada}

\author{Eduardo Mart\'{i}n-Mart\'{i}nez}
\email{emartinmartinez@uwaterloo.ca}
\affiliation{Department of Applied Mathematics, University of Waterloo, Waterloo, Ontario, N2L 3G1, Canada}
\affiliation{Institute for Quantum Computing, University of Waterloo, Waterloo, Ontario, N2L 3G1, Canada}
\affiliation{Perimeter Institute for Theoretical Physics, Waterloo, Ontario, N2L 2Y5, Canada}

\begin{abstract}
  
  We analyzed how entanglement harvesting is affected by the performance of a measurement on the quantum field. The measurement on the field is modelled as the coupling of a particle detector to the field, followed by a projective measurement performed on the detector. In our analysis, we considered different arrangements for two detectors harvesting entanglement and an ancillary detector used to perform a measurement. We found different regimes for how performing measurements on the field affects the entanglement harvested, depending on the initial and final states of the detector used to measure the field, as well as its coupling strength. We identified the regimes where it is possible to measure the field during the preparation of entanglement harvesting protocols without significantly interfering in the entanglement harvested. We also identify in what regimes the field measurement can degrade or slightly enhance the ability of particle detectors to harvest entanglement. 
  
\end{abstract}

\maketitle

\section{Introduction}

The entanglement structure of quantum field theories (QFTs) plays a relevant role in a variety of phenomena, ranging from quantum energy teleportation~\cite{Hotta2009,Hotta2011} and the black hole information loss problem~\cite{Preskill1992}, to the probing of the confinement transition phase in quantum chromodynamics~\cite{Klebanov2008,Jokela2021}, and the definition of a measurement theory for quantum fields~\cite{ReehSchlieder,Schlieder1968,Sorkin,Borsten2021}. It is well-known that quantum fields generically display classical and quantum correlations between arbitrary regions of spacetime~\cite{SUMMERS, vacuumEntanglement}. In fact, the ultraviolet divergence of entanglement entropy~\cite{Bombelli1986} shows that this is a characteristic feature of the algebra of observables itself, rather than just a feature of the state of the quantum field~\cite{witten}.

This entanglement can in fact be extracted by pairs of particle detectors, i.e., non-relativistic quantum probes which couple locally to the field. The entanglement acquired by the particle detectors witnesses the entanglement displayed by the field between the interaction regions. 
This phenomena, first noted in~\cite{Valentini1991} and later rediscovered in~\cite{Reznik2003,Reznik1,Reznik2}, has become known as \textit{entanglement harvesting}~\cite{Salton:2014jaa} and has been the subject of exhaustive study (see, among many others,~\cite{Nick,Cliche2011,Pozas-Kerstjens:2015,Pozas2016,topology,PetarOld,Petar,Ng2,Henderson2018,Henderson2019,Henderson2020,Erickson2020,ericksonNew,Foo2021,carol,Sahu2022,Bueley2022,MendezAvalos2022}). Apart from providing an operational way of accessing the entanglement of quantum fields, entanglement harvesting is a proxy witness for the geometry~\cite{Nick,Ng2,Henderson2019} and topology~\cite{topology} of spacetime.

One important feature of particle detectors is that they connect with actual experimental realizations. In particular, the Unruh-DeWitt (UDW) particle detector~\cite{Unruh1976,DeWitt} captures the most relevant features of the interaction between atoms and the electromagnetic field~\cite{eduardoOld,eduardo,richard}. With the actual technology in hand, entanglement harvesting should in principle be feasible to reproduce in experimental setups~\cite{Sabin2010,Sabin2012,Borrelli2012,FornDiaz2017,Ardenghi2018,Janzen2022}. 

Measurement and state preparation are intrinsic parts of any experimental setup in physics. In order to faithfully model realistic entanglement harvesting protocols we need to know the effect of measurements on the harvested correlations. For instance, the usual setups discussed in the literature often assume the field to be initialized in its vacuum state (see, e.g.,~\cite{Salton2015,Pozas-Kerstjens:2015,Zhang2020}, among many others) or in coherent states~\cite{PetarOld,Petar}. Hence, any realistic implementation of these protocols requires a fair understanding of how the very preparation of the setup or even measurements from previous experiments affect our results. 

Since measurements are an intrinsic part of any experimental setup in physics, in order to approach realistic scenarios in which entanglement harvesting protocols might be implemented we need to analyze the effect of such measurements in the harvested correlations. For instance, the usual setups discussed in the literature often assume the field to be initialized in its vacuum state (see, e.g.,~\cite{Salton2015,Pozas-Kerstjens:2015,Zhang2020}, among many others) or in coherent states~\cite{PetarOld,Petar}. Hence, any realistic implementation of these protocols requires a fair understanding of how the very preparation of the setup or even measurements from previous experiments affect our results. 

The issue of how to model measurements on quantum fields is an open problem since Rafael Sorkin pointed out in 1992 that the projection postulate could not be exported from non-relativistic quantum mechanics to quantum field theory~\cite{Sorkin,Borsten2021}. There are currently two main approaches, which differ in the way they model the probe with which the measurement is performed: either treating the probe as a quantum field~\cite{fewster1,fewster2,fewster3} or as a particle detector~\cite{Jose}. Here, we will use the second approach, where the measurement process is modelled using a particle detector that is first let to interact with the field, thus gathering information from the field through the interaction. After the detector is decoupled from the field, a projective measurement is performed on the detector, allowing the experimenter to extract information from the detector, and thus about the field.

In this paper we aim to analyze the protocol of entanglement harvesting by a pair of particle detectors in the presence of a measurement performed by a third party, who operates another particle detector. We will study how the result of the measurement and the propagation of the information about the measurement outcome influences our ability to harvest entanglement. In Section~\ref{Setup} we introduce the setup considered with two particle detectors harvesting entanglement from a quantum field and an additional detector used to measure it. In Section~\ref{regimes} we identify different regimes in perturbation theory depending on the relations between the parameters of the problem and quantify the entanglement harvested in each regime. In Section~\ref{explicitProtocol} we consider specific protocols with Gaussian switchings and pointlike smearings in order to explicitly calculate the effect of the measurement on the entanglement in the regimes in which it modifies it at leading order. In Section~\ref{Non-selective} we analyze the effect of non-selective measurements. The conclusions of this manuscript are presented in Section~\ref{Conclusions}.


\section{Setup} \label{Setup}
In this work we consider three experimenters, Alba, Blanca, and Clara, undergoing timelike trajectories parametrized by their proper times as $\mf{z}_{\ts a}(\tau_{\ts{a}})$, $\mf{z}_{\ts b}(\tau_{\ts{b}})$, and $\mf{z}_{\ts c}(\tau_{\ts{c}})$.
Each experimenter is equipped with a probe that we model as a spatially smeared Unruh-DeWitt particle detector~\cite{Unruh1976,DeWitt}, weakly coupled to a real massless scalar field $\phih(\mf x)$ in a \mbox{$(3+1)$-dimensional} Minkowski spacetime. Each particle detector is modelled as a two-level quantum system with a proper energy gap $\Omega_{\ts i}$ between  the ground state $\ket{g_{\ts i}}$ and the excited state $\ket{e_{\ts i}}$, for $\ts I \in \{\ts A, \ts B, \ts C\}$. The particle detectors are initialized in their ground states, i.e., the density operator that describes their initial joint state reads
\begin{align}
    \rhoh_{\ts{abc}}^{(0)} = \proj{g_{\ts a}}{g_{\ts a}} \otimes \proj{g_{\ts b}}{g_{\ts b}} \otimes \proj{g_{\ts c}}{g_{\ts c}}.
\end{align}
The scalar field $\phih(\mathsf{x})$ can be expanded in plane-wave modes as
\begin{equation}
    \hat{\phi}(\mf x) = \int \frac{\dd^3 \bm{k}}{\sqrt{(2\pi)^3 2|\bm{k}|}} \,(\hat{a}_{\bm k}\,e^{-\ii(|\bm{k}|t - \bm{k} x)} + \text{H.c.}),
\end{equation}
where $(t,\bm{x})$ are some inertial coordinates for the event $\mf{x}$, and $\hat{a}_{\bm k}$ and $\hat{a}_{\bm k}^{\dagger}$ are the annihilation and creation operators satisfying the canonical commutation relations \mbox{$[\hat{a}_{\bm k}, \hat{a}_{\bm k'}^{\dagger}] = \delta^{(3)}(\bm{k} - \bm{k}')$}. We consider the field to be initially in the vacuum state $\rhoh^{(0)}_{\phi} = \proj{0}{0}$ which satisfies $\hat{a}_{\bm k}\ket{0} = 0$ for all $\bm k$. Thus, the initial joint state of the detectors-field system is
\begin{equation}
    \rhoh^{(0)}= \rhoh_{\ts{abc}}^{(0)} \otimes \proj{0}{0}.
\end{equation}
The particle detectors couple locally to the quantum field according to the  Hamiltonian weight~\cite{us}
\begin{equation} \label{eq:hamiltonianW}
    \hat{h}(\mf x) = \hat{h}_{\ts{a}}(\mf x) +  \hat{h}_{\ts{b}}(\mf x) + \hat{h}_{\ts{c}}(\mf x),
\end{equation}
where the single interaction between each detector and the field is given in the interaction picture by
\begin{equation}
    \hat{h}_{\ts j}(\mf x) = \lambda_{\ts j} \Lambda(\mf x) \muh(\tau_{\ts j}) \hat{\phi}(\mf x).
\end{equation}
Here, $\lambda_\textsc{j}$ is the coupling strength, the spacetime smearing $\Lambda_\textsc{j}(\mf{x})$ determines the relative strength of the interaction between the particle detector and the quantum field at $\mf{x}$, and $\muh_{\ts j}(\tau_{\ts j})$ is the monopole moment operator of each particle detector, which acts over the energy eigenbasis as
\begin{align}
    \muh_{\ts j}(\tau_{\ts j}) \ket{g_{\ts j}} &= e^{\ii \Omega_{\ts j} \tau_{\ts j}} \ket{e_{\ts j}},\\
    \muh_{\ts j}(\tau_{\ts j}) \ket{e_{\ts j}} &= e^{-\ii \Omega_{\ts j} \tau_{\ts j}} \ket{g_{\ts j}}.
\end{align}
The time-evolution operator that implements the unitary evolution due to the Hamiltonian weight in Eq.~\eqref{eq:hamiltonianW} is then given by
\begin{equation}
    \hat{U} = \mathcal{T}\text{exp}\left(-\ii \int \dd V\hat{h}(\mf x) \right),
\end{equation}
where $\mathcal{T}$ denotes time-ordering with respect to an arbitrary time-parameter $t$, and $\dd V$ is the invariant volume measure over spacetime. In the regimes that we consider here, the predictions of the model are independent of the choice of this time parameter~\cite{us2}. The time-evolution operator maps the joint state of the detectors and the field before the interaction to the joint state after the interaction $\rhoh = \hat{U} \, \rhoh^{(0)} \hat{U}^{\dagger}$.

To incorporate the effect of measurements in this setup, once Clara's detector interaction is switched off, she measures a certain observable of her particle detector. Following the formalism in~\cite{Jose}, this process is modelled applying a rank-one projector \mbox{$\hat{P} = \proj{s}{s}$} on the detector's state, corresponding to the outcome $s$ of the measurement. Without loss of generality, we can write the outcome state $\ket{s}$ as\footnote{Since we are working in the interaction picture, the specific form of the projector $\hat{P}$ depends on the time in which the measurement is performed, since the outcome state $\ket{s}$ has to be evolved with the free Hamiltonian correspondingly. This in turn means that the phase $\xi$ in Eq.~\eqref{eq:outcome s} depends on the time of the performance of the measurement. For the sake of a simpler notation, we will ignore this subtlety in our calculations, compensating for it by studying the dependence on the phase $\xi$ later on.}
\begin{equation}\label{eq:outcome s}
    \ket{s} = \epsilon\ket{g_{\ts c}} + \sqrt{1-\epsilon^2}\,e^{\ii \xi} \ket{e_{\ts c}},
\end{equation}
where $\epsilon$ is a non-negative real number, and $\xi \in [0,2\pi)$. Unlike performing projective measurements directly on the field~\cite{Sorkin,Borsten2021}, this measurement is compatible with relativistic causality~\cite{Jose}. 
Note that the projector $\hat{P}$ can always be applied after the interactions of detectors A and B with the field have finished, since $\hat{P}$ commutes with the Hamiltonian weights $\hat{h}_{\ts a}(\mf x)$ and $\hat{h}_{\ts b}(\mf x)$. This is true even if the measurement is performed on detector C while detectors A and B are still coupled to the field. The joint detectors-field system (including the time-evolution and the measurement) is then
\begin{equation} \label{eq:finalJointState}
    \rhoh^{s} = \frac{\hat{P}\hat{U} \rhoh^{(0)}\hat{U}^{\dagger} \hat{P}}{\text{Tr}\big(\hat{P}\hat{U} \rhoh^{(0)}\hat{U}^{\dagger} \hat{P} \big)}.
\end{equation}
The state above represents a selective update, corresponding to an observer who has access to the outcome of the measurement performed on C---and therefore has to be placed in the causal future of the measurement performance~\cite{Jose}. The goal is to analyze the entanglement harvested by detectors A and B in order to understand the effect that the measurement had on it. To do so, we trace out the degrees of freedom corresponding to the field and to detector C, yielding
\begin{align}
    \rhoh^{s}_{\ts{ab}} &= \frac{\text{Tr}_{\ts c, \phi}\big(\hat{P}\hat{U} \rhoh^{(0)}\hat{U}^{\dagger} \hat{P}\big)}{\text{Tr}\big(\hat{P}\hat{U} \rhoh^{(0)}\hat{U}^{\dagger} \hat{P} \big)}= \frac{\bra{s}\!\text{Tr}_{\phi}\big(\hat{U} \rhoh^{(0)}\hat{U}^{\dagger} \big)\!\ket{s}}{\text{Tr}\big(\hat{P}\hat{U} \rhoh^{(0)}\hat{U}^{\dagger} \hat{P} \big)} \;. 
    \label{eq:selective}
\end{align}
The time-evolved state does not admit a closed form, but working in a weak coupling regime allows us to treat it perturbatively in the coupling strengths. In order to perform the perturbative analysis in a consistent way, we need to specify how the orders of the different parameters are related. Here, we assume that all the coupling strengths are equal, \mbox{$\lambda_{\ts{a}} = \lambda_{\ts{b}} = \lambda_{\ts{c}} \equiv \lambda$}. This assumption allows us to write the perturbative series in terms of a single coupling strength $\lambda$, simplifying the analysis significantly\footnote{In particular, doing this we can jointly analyze the measurement and the harvesting protocol just considering the leading order of the perturbative series. One could indeed analyze the case where $\lambda_{\ts{a}}$, $\lambda_{\ts{b}}$, and $\lambda_{\ts{c}}$ are of the same order of magnitude but different from each other using the same procedure. In this case, since the magnitudes of the three strengths are similar, the scale that characterizes the different orders in perturbation theory would be the one that sets the magnitude of the three couplings. However, in this paper we restrict ourselves to identical coupling strengths to avoid the increased calculational overhead that such an analysis would require.

Other cases would in general require considering higher orders to see the interplay between the measurement and the protocol. For instance, if \mbox{$\lambda_{\ts{a}},\lambda_{\ts{b}}\sim o(\lambda_{\ts{c}})$}, then the effect of the measurement on the field state would be predominant, and we would have to go further than leading order to see how the measurement affects the (now weaker) harvesting protocol. On the contrary, if \mbox{$\lambda_{\ts{c}}^2\sim o(\lambda_{\ts{a}}\lambda_{\ts{b}})$}, then the harvesting protocol would dominate, and we would have to consider higher order terms to analyze the effect of the measurement.}. The Dyson expansion of the time evolution operator then reads
\begin{equation}
    \hat{U} = \mathds{1} + \sum_{k \geq 1} \hat{U}^{(k)},
\end{equation}
with 
\begin{align}
    \hat{U}^{(k)} &= (-\ii)^k\int \dd V_1 \ldots \dd V_k \, \,\hat{h}(\mf x_1) \ldots \hat{h}(\mf x_k)\nonumber \\&\phantom{=========} \times\theta(t_1 - t_2) \ldots \theta(t_{k-1} - t_k)\label{eq:Dyson}.
\end{align}
In particular, the numerator of Eq. \eqref{eq:selective} can be expanded as (see Appendix~\ref{appendix numerator} for details)
\begin{align} \label{eq:numerator}
    \text{Tr}_{\ts c, \phi}\big(\hat{P}\hat{U} \rhoh^{(0)}\hat{U}^{\dagger} \hat{P}\big) &= \epsilon^2 (\rhoh_{\ts{ab}}^{(0)} + \rhoh_{\ts{ab}}^{(2)}) \nonumber \\ 
    &\phantom{=\,}+ \left(\mathcal{L}_{\ts{cc}}(1-2\epsilon^2) + \kappa\right)\rhoh_{\ts{ab}}^{(0)}  \nonumber \\
    &\phantom{=\,}+\hat{\gamma} + \hat{\nu} + \epsilon\,\mathcal{O}(\lambda^4) + \mathcal{O}(\lambda^6),
\end{align}
where the expression for $\kappa$ is obtained in Appendix~\ref{appendix numerator}, and given in Eq.~\eqref{eq:kappa}. In the basis $\{\ket{g_\textsc{a} g_{\textsc{b}}}, \ket{g_\textsc{a} e_{\textsc{b}}}, \ket{e_\textsc{a} g_{\textsc{b}}}, \ket{e_\textsc{a} e_{\textsc{b}}}\}$, the operators above can be written in matrix form as
\begin{equation}
    \rhoh^{(0)}_{\ts{ab}} = \left(
\begin{array}{cccc}
1 & 0 & 0 & 0 \\
0 & 0 & 0 & 0 \\
0 & 0 & 0 & 0 \\
0 & 0 & 0 & 0 \\
\end{array}\right), 
\end{equation}
\begin{equation}
    \rhoh^{(2)}_{\ts{ab}} = \left(
\begin{array}{cccc}
-\mathcal{L}_\textsc{aa}-\mathcal{L}_\textsc{bb} & 0 & 0 & \mathcal{M}_{\ts{ab}}^\ast \\
0 & \mathcal{L}_\textsc{bb} & \mathcal{L}_\textsc{ab}^\ast & 0 \\
0 & \mathcal{L}_\textsc{ab} & \mathcal{L}_\textsc{aa} & 0 \\
\mathcal{M}_{\ts{ab}} & 0 & 0 & 0 \\
\end{array}\right), \label{eq:rho2}
\end{equation}
\begin{equation}
    \hat{\gamma} = \epsilon\sqrt{1-\epsilon^2}\left(
\begin{array}{cccc}
0 & 0 & 0 & 0 \\
e^{\ii \xi}\mathcal{M}_{\ts{ac}} + e^{-\ii \xi}\mathcal{L}_{\ts{ac}} & 0 & 0 & 0 \\
e^{\ii \xi}\mathcal{M}_{\ts{bc}} + e^{-\ii \xi}\mathcal{L}_{\ts{bc}} & 0 & 0 & 0 \\
0 & 0 & 0 & 0 \\
\end{array}\right) + \text{H.c.},   \label{eq:gamma}
\end{equation}
\begin{equation}
    \hat{\nu} = \mathcal{L}_{\ts{cc}}\left(
\begin{array}{cccc}
-\Tilde{\mathcal{L}}_\textsc{aa}-\Tilde{\mathcal{L}}_\textsc{bb} & 0 & 0 & \Tilde{\mathcal{M}}_{\ts{ab}}^\ast \\
0 & \Tilde{\mathcal{L}}_\textsc{bb} & \Tilde{\mathcal{L}}_\textsc{ab}^\ast & 0 \\
0 & \Tilde{\mathcal{L}}_\textsc{ab} & \Tilde{\mathcal{L}}_\textsc{aa} & 0 \\
\Tilde{\mathcal{M}}_{\ts{ab}} & 0 & 0 & 0 \\
\end{array}\right) \label{eq:nu}.
\end{equation}
Here, for \mbox{$\ts{I,\,J} \in \{ \ts{A},\ts{B}, \ts{C}\}$},
\begin{align}
    \mathcal{L}_{\ts{ij}} &= \lambda^2 \int \dd V \dd V' \Lambda_{\ts i}(\mathsf{x}) \Lambda_{\ts j}(\mathsf{x}') \, e^{\ii(\Omega_{\ts i} \tau_{\ts i} - \Omega_{\ts j} \tau'_{\ts j})} \, W(\mf x', \mf x). \label{eq:Lij}
\end{align}
Also, for $\ts{I} \neq \ts{J}$,
\begin{align}
    \mathcal{M}_{\ts{ij}} &= -\lambda^2 \int \dd V \dd V'\, \Lambda_{\textsc{i}}(\mathsf{x}) \Lambda_{\textsc{j}}(\mathsf{x}')\, e^{\ii(\Omega_{\textsc{i}} \tau_\ts{i} + \Omega_{\textsc{j}} \tau'_\textsc{j})} \nonumber \\  &\phantom{=}\times\left( \theta(t-t')W(\mf x, \mf x')  +\theta(t'-t)W(\mf x', \mf x) \right), \label{eq:Mij}
\end{align}
and for $\ts{I,\,J} \in \{ \ts{A},\ts{B}\}$,
\begin{align}
    \Tilde{\mathcal{L}}_{\ts{ij}} &= \mathcal{L}_{\ts{ij}} + \frac{\mathcal{L}_{\ts{ic}} \mathcal{L}_{\ts{jc}}^{*} + \mathcal{M}_{\ts{ic}} \mathcal{M}_{\ts{jc}}^{*}}{\mathcal{L}_{\ts{cc}}} \;, \label{eq:TLij}\\
    \Tilde{\mathcal{M}}_{\ts{ab}} &=  \mathcal{M}_{\ts{ab}} + \frac{\mathcal{L}_{\ts{ac}} \mathcal{M}_{\ts{bc}} + \mathcal{L}_{\ts{bc}} \mathcal{M}_{\ts{ac}}}{\mathcal{L}_{\ts{cc}}} \label{eq:TMij},
\end{align}
where \mbox{$W(\mf x, \mf x') = \bra{0} \phih(\mf x) \phi(\mf x') \ket{0}$} is the field Wightman function. Note that $\rhoh_{\ts{ab}}^{(0)}=\proj{g_{\ts a}}{g_{\ts a}} \otimes \proj{g_{\ts b}}{g_{\ts b}}$ is simply the initial joint partial state of detectors A and B. The terms $\rhoh_{\ts{ab}}^{(2)}$ and $\hat{\gamma}$ are of second order in $\lambda$, whereas $\hat{\nu}$ and $\kappa$ are of fourth order in $\lambda$. In particular, $\kappa$ is subleading for all the calculations of interest. Finally, in Eq.~\eqref{eq:selective}, the denominator is simply the trace of the numerator, which, using Eq.~\eqref{eq:numerator}, we can write as
\begin{align} \label{eq:denominator}
    \text{Tr}\big(\hat{P}\hat{U} \rhoh^{(0)}\hat{U}^{\dagger} \hat{P} \big) &= \epsilon^2 + \mathcal{L}_{\ts{cc}}(1-2\epsilon^2) \nonumber \\
    &\phantom{===}+ \kappa + \epsilon\, \mathcal{O}(\lambda^4) + \mathcal{O}(\lambda^6) \;.
\end{align}
This denominator is in fact the probability $\textrm{Prob}(s)$ of measuring the outcome $\ket{s}$. Along with Eq.~\eqref{eq:numerator}, this allows us to find a perturbative expansion for the selectively updated joint density operator $\rhoh_{\ts{abc}}^{s}$ given in Eq.~\eqref{eq:selective}. As we will discuss later, it suffices to perform this expansion up to second order in perturbation theory, since this is the leading order for the entanglement harvested. This involves considering a perturbative expansion of Eq.~\eqref{eq:selective}. Note, however, that the expression for $\rhoh_{\ts{ab}}^{s}$ is not analytic at $(\epsilon, \lambda) = (0,0)$
. This forces us to be specially careful with the expansions when $\epsilon = \braket{s}{g_{\ts c}}$ approaches zero (i.e., when the outcome state $\ket{s}$ is close to the excited state $\ket{e_\textsc{c}}$). The necessity of studying separately the case in which the outcome of the measurement is orthogonal to the initial state of the measured particle detector was already pointed out in~\cite{Jose}. One way of making sense of this is to realize that one cannot expect the updated state of the field for $\epsilon=0$ to converge to the vacuum when $\lambda_{\ts c} \to 0$, since if the particle detector does not couple to the field at all, then an orthogonal outcome is not possible. However, it is important to analyze the case where $1\gg\epsilon>0$, for which we can take a perturbative approach on the parameter $\epsilon$. 

In order to gain further intuition, let us analyze the behaviour of $\textrm{Prob}(s)$, given in Eq.~\eqref{eq:denominator}, at leading order. If $\lambda \ll \epsilon$, i.e., if the inner product $|\!\braket{s}{g_\ts{c}}\!|$ is large compared to the coupling strength\footnote{Notice that we are considering that the background spacetime of our setup is $(3+1)$-dimensional. Unlike in any other number of dimensions, for 3+1 spacetime dimensions the coupling strength $\lambda$ is dimensionless. Therefore, its magnitude can be directly compared with the inner product $\epsilon = |\!\braket{s}{g_{\ts c}}\!|$, which is also a dimensionless quantity.}, $\textrm{Prob}(s)$ is given at leading order by $\epsilon^2$. In particular, in Eq.~\eqref{eq:denominator}, the term proportional to $\epsilon^2$ is one order of $\lambda$ below the rest of the terms as long as $\epsilon = \Theta(\lambda^\delta)$ for $\delta < 1/2$, where we have used the big theta symbol\footnote{Specifically, we say that $f(x)=\Theta(g(x))$ when $x \to a$ (for $a \in \bar{\mathbb{R}}$) iff there exist positive constants $0<c_1<c_2$ such that
\begin{equation*}
  c_1 < \liminf_{x\to a}\frac{f(x)}{g(x)} \leq \limsup_{x\to a}\frac{f(x)}{g(x)} <c_2 . 
\end{equation*}
This is in contrast with the big O (or big Omicron) notation, $f(x)=\mathcal{O}(g(x))$, which shall be interpreted as
$    \limsup_{x \to a} \left|f(x)/g(x)\right| < \infty \;.
$} 
of the Bachmann-Landau notation~\cite{Knuth1976}. We will refer to the regime $\epsilon=\Theta(\lambda^\delta)$ for $\delta < 1/2$ as the \textit{non-orthogonal regime}. This is because in the perturbative regime $\lambda \ll 1$, $\epsilon=\Theta(\lambda^\delta)$ with $\delta < 1/2$ guarantees that $\epsilon \gg \lambda$ (roughly speaking, the distance to orthogonality is much larger than the displacement that the interaction will cause)
. If $\epsilon = \Theta(\lambda^\delta)$ with $\delta > 3/2$, the term proportional to $\epsilon^2$ is instead one order in $\lambda$ above the leading order in Eq.~\eqref{eq:denominator}, which is given by the excitation probability $\mathcal{L}_{\ts{cc}}=\Theta(\lambda^2)$. We will call this regime the \textit{orthogonal regime}, since in this case $\epsilon \ll \lambda \ll 1$. Finally, if $\epsilon= \Theta(\lambda^\delta)$ for $1/2<\delta<3/2$, the leading order term in $\lambda$ in Eq.~\eqref{eq:denominator} is $\epsilon^2 + \mathcal{L}_{\ts{cc}}$. We will refer to this regime as the \textit{transition regime}.
We will study the entanglement harvested for the different regimes separately
. To understand the effect that the measurement has in each case, we shall compare the results with the entanglement that Alba and Blanca would harvest if Clara were not there to perform her measurement. In the absence of measurements, the time evolved state would be
\begin{equation}
    \rhoh_{\ts{ab}} = \text{Tr}_{\phi}\left(\hat{V} \rhoh_{\ts{ab}}^{(0)} \hat{V}^{\dagger}\right),
\end{equation}
where $\hat{V}$ is the time-evolution operator associated only to the interactions of detectors A and B, namely
\begin{equation}
    \hat{V} = \mathcal{T}\text{exp}\left(-\ii \int \dd V(\hat{h}_{\ts a}(\mf x) +  \hat{h}_{\ts b}(\mf x))\right).
\end{equation}
With the notation used in Eq.~\eqref{eq:numerator}, the final state is simply~\cite{Pozas-Kerstjens:2015}
\begin{equation}\label{eq:state wo Clara}
    \rhoh_{\ts{ab}} = \rhoh_{\ts{ab}}^{(0)} + \rhoh_{\ts{ab}}^{(2)} + \mathcal{O}(\lambda^4).
\end{equation}

\section{Entanglement harvesting when the field is measured} \label{regimes}
In order to quantify the entanglement between Alba and Blanca's particle detectors, we use the negativity~\cite{Nielsen2010}. We will compute the negativity of $\rhoh^{s}_{\ts{ab}}$ in each regime and compare it with the negativity of $\rhoh_{\ts{ab}}$ corresponding to the protocol without measurements (wm), which reads~\cite{Pozas-Kerstjens:2015}
\begin{align}
    \mathcal{N}_{\textrm{wm}}&= \max\left(\!0\,,\sqrt{|\mathcal{M}_{\ts{ab}}|^2\! +\!\frac{(\L_{\ts{aa}}\! -\! \L_{\ts{bb}})^2}{4}}\! -\!\frac{\L_{\ts{aa}}\! +\!\L_{\ts{bb}}}{2} \right) \nonumber\\
    &\phantom{=\,}+\mathcal{O}(\lambda^4) \label{eq:negativity0} \;.
\end{align}
We can associate the term $\mathcal{M}_{\ts{ab}}$ with the quantum correlations acquired by detectors A and B. $\mathcal{L}_{\ts{aa}}$ and $\mathcal{L}_{\ts{bb}}$ are local terms corresponding to the excitation probabilities of each detector. Note that the expression is clearly bounded by $|\mathcal{M}_{\ts{ab}}|$, and the local terms act as noise that can only decrease the entanglement harvested by the detectors.

If, instead, we consider a protocol in which a third party has performed a measurement, the resulting post-measurement state has the following structure in matrix form:
\begin{align} \label{eq:matrix1}
     \rhoh^{s}_{\ts{ab}}=&\left(
\begin{array}{cccc}
1-\lambda^2r_{22}-\lambda^2r_{33} & \Theta(\lambda^b) & \Theta(\lambda^b) & \lambda^2r_{41}^\ast \\
\Theta(\lambda^b) & \lambda^2 r_{22} & \lambda^2 r_{32}^* & 0 \\
\Theta(\lambda^b) & \lambda^2 r_{32} & \lambda^2 r_{33} & 0 \\
\lambda^{2} r_{41} & 0 & 0 & 0 \\
\end{array}\right) \nonumber\\
&+\mathcal{O}(\lambda^3) \;,
\end{align}
where $r_{22}$ and $r_{33}$ are positive and $b\geq 1$. In the case $b \geq 3/2$, which corresponds to both the orthogonal and the non-orthogonal regime, the negativity is given by
\begin{align} \label{eq:negativity1}
    \mathcal{N}_s&= \lambda^2\,\max\left(0\,,\sqrt{|r_{41}|^2+\frac{(r_{22} - r_{33})^2}{4}} - \frac{r_{22} + r_{33}}{2}\right) \nonumber \\ &\phantom{=\,}+ \mathcal{O}(\lambda^{3}),
\end{align}
as seen in Appendix \ref{appendix negativity}. The cases with $1 \leq b < 3/2$ correspond to the transition regime. In what follows, we will analyze the non-orthogonal and orthogonal regimes in full generality, and  analyze the particular case of the transition regime in which \mbox{$b = 1$}, for which the negativity admits a second order power expansion in $\lambda$.

\subsection{Non-orthogonal regime}\label{Non-orthogonal regime}
We analyze the case in which the scalar product between the initial state of detector C, $\ket{g_{\ts c}}$, and the outcome of the measurement, $\ket{s}$, are far from being orthogonal. This translates into the assumption \mbox{$\epsilon = \braket{s}{g_{\ts c}} = \Theta(\lambda^\delta)$}, with $\delta < 1/2$. Under these conditions, the probability of the outcome $\ket{s}$ is at leading order $\epsilon^2$, and since $\lambda \ll 1$, we have that $\epsilon \gg \lambda$. Note that by setting $\delta=0$ we recover the case in which $\epsilon$ is just some finite constant unrelated to the coupling between the field and detector C. Under these conditions, the state updated after the measurement can be expanded as
\begin{align}
    &\rhoh_{\ts{ab}}^{s} = \rhoh_{\ts{ab}}^{(0)} +  \frac{1}{\epsilon^2}\hat{\gamma} + \rhoh_{\ts{ab}}^{(2)} - \frac{\mathcal{L}_{\ts{cc}}}{\epsilon^4}\hat{\gamma} + \mathcal{O}(\lambda^3)  \\
    &\hspace{-0.1cm}= \left(
\begin{array}{cccc}
1-\mathcal{L}_\textsc{aa}-\mathcal{L}_\textsc{bb} & \Theta(\lambda^{2-\delta}) & \Theta(\lambda^{2-\delta}) & \mathcal{M}_{\ts{ab}}^\ast \\
\Theta(\lambda^{2-\delta}) & \mathcal{L}_\textsc{bb} & \mathcal{L}_\textsc{ab}^\ast & 0 \\
\Theta(\lambda^{2-\delta}) & \mathcal{L}_\textsc{ab} & \mathcal{L}_\textsc{aa} & 0 \\
\mathcal{M}_{\ts{ab}} & 0 & 0 & 0 \\
\end{array}\right) + \mathcal{O}(\lambda^3). \nonumber
\end{align}
All the measurement contributions are proportional to the operator $\hat{\gamma}$, and they are contained in the $\Theta(\lambda^{2-\delta})$ entries of the measurement updated final operator.  Using the expression given in Eq. \eqref{eq:negativity1}, we find that the negativity for the measurement updated density operator is given by
\begin{align}
    \mathcal{N}_s&= \max\left(\!0\,,\sqrt{|\mathcal{M}_{\ts{ab}}|^2\! +\!\frac{(\L_{\ts{aa}}\! -\! \L_{\ts{bb}})^2}{4}}\! -\!\frac{\L_{\ts{aa}}\! +\!\L_{\ts{bb}}}{2} \right) \nonumber\\
    &\phantom{=\,}+\mathcal{O}(\lambda^3), \;
\end{align}
which coincides at leading order in $\lambda$ with the negativity in the absence of measurements, given in Eq.~\eqref{eq:negativity0}. Despite the measurement contributing to the joint state of the particle detectors at leading order, the negativity at leading order is fully determined by the interactions of detectors A and B, and therefore independent of the measurement. 

Experimental implementations of entanglement harvesting might imply measuring the quantum field during the preparation of the protocol
. These results imply that within the non-orthogonal regime it is possible to perform \textit{weak} measurements on the quantum field without significantly affecting the entanglement harvested in an independent protocol. One easy way to guarantee this is to only measure the particle detector used to probe the field in a basis whose elements are not close to orthogonal to the initial state of such detector (which in this case is its ground state).

\subsection{Orthogonal regime} \label{Orhtogonal regime}
We now focus on the case in which the scalar product is small compared to the coupling parameter, i.e. \mbox{$\epsilon \ll \lambda \ll 1$}. Specifically, we consider $\epsilon = \Theta(\lambda^{\delta})$ with $\delta > 3/2$. In this regime, the probe's post-measurement state $\ket{s}$ is close enough to the excited state so that the probability of obtaining that outcome, $\textrm{Prob}(s)$, is dominated by the probability of excitation of detector C, $\mathcal{L}_{\ts{cc}}$. Under these conditions, the joint state of the detectors can be expressed as
\begin{align}
    &\rhoh_{\ts{ab}}^{s}= \rhoh_{\ts{ab}}^{(0)} + \frac{\hat{\gamma}}{\mathcal{L}_{\ts{cc}}}+ \frac{\hat{\nu}}{\mathcal{L}_{\ts{cc}}} + \mathcal{O}(\lambda^3) \\
    &= \left(
\begin{array}{cccc}
1-\Tilde{\mathcal{L}}_\textsc{aa}-\Tilde{\mathcal{L}}_\textsc{bb} & \Theta(\lambda^\delta) & \Theta(\lambda^\delta) & \Tilde{\mathcal{M}}_{\ts{ab}}^\ast \\
\Theta(\lambda^\delta) & \Tilde{\mathcal{L}}_\textsc{bb} & \Tilde{\mathcal{L}}_\textsc{ab}^\ast & 0 \\
\Theta(\lambda^\delta)& \Tilde{\mathcal{L}}_\textsc{ab} & \Tilde{\mathcal{L}}_\textsc{aa} & 0 \\
\Tilde{\mathcal{M}}_{\ts{ab}} & 0 & 0 & 0 \\
\end{array}\right) + \mathcal{O}(\lambda^3). \nonumber
\end{align} 
Since $\delta > 3/2$, we can again use Eq. \eqref{eq:negativity1} to compute the negativity:
\begin{align} 
    \mathcal{N}_s&= \text{max}\left(\!0,\sqrt{|\tilde{\mathcal{M}}_{\ts{ab}}|^2\! +\!\frac{(\Tilde{\L}_{\ts{aa}}\! -\! \Tilde{\L}_{\ts{bb}})^2}{4}}\! -\!\frac{\Tilde{\L}_{\ts{aa}}\! +\!\Tilde{\L}_{\ts{bb}}}{2} \right) \nonumber\\
    &\phantom{=\,}+\mathcal{O}(\lambda^3). \label{eq:NegativityOrthogonal}\;
\end{align}
In this case, the measurement performed on C does modify the leading order of the entanglement harvested by A and B. On the one hand, from Eq.~\eqref{eq:Lij} we see that the measurement enhances the local noise terms by adding positive contributions of the form \mbox{$(|\mathcal{L}_{\ts{ic}}|^2 + |\mathcal{M}_{\ts{ic}}|^2)/\mathcal{L}_{\ts{cc}}$}, for $\ts{I} \in \{\ts{A}, \ts{B} \}$. On the other hand, from Eq.~\eqref{eq:Mij} we also see that the measurement modifies the non-local correlation term. Since the entanglement harvested is a competition between local and non-local terms~\cite{Pozas-Kerstjens:2015}, it is not clear from the equation above whether the measurement increases or decreases the entanglement harvested by the detectors. In Section~\ref{explicitProtocol}, we will restrict ourselves to a specific case study in order to obtain some explicit results.

\subsection{Transition regime}
\label{Transition regime}

We now explore the case where $\epsilon = \Theta(\lambda^\delta)$ with \mbox{$1/2 < \delta < 3/2$}. Specifically, we will analyze the case where  $\epsilon$ and $\lambda$ are comparable, i.e., when $\delta = 1$. This is convenient because for the remaining range of values of $\delta$, the negativity does not admit a power series up to second order in $\lambda$. In the case $\delta=1$, the final state of the particle detectors can be expanded as
\begin{align}
    \rhoh_{\ts{ab}}^s &=  \rhoh_{\ts{ab}}^{(0)} + \frac{\hat{\gamma} + \hat{\nu} + \epsilon^2\rhoh_{\ts{ab}}^{(2)}}{\epsilon^2 + \mathcal{L}_{\ts{cc}}} + \mathcal{O}(\lambda^3)  \label{eq:matrixTransition}\\
 &= \left(
\begin{array}{cccc}
1-\mathcal{L}_\textsc{aa}'-\mathcal{L}_\textsc{bb}' & \mathcal{L}_\textsc{b}'^* & \mathcal{L}_\textsc{a}'^*  & \mathcal{M}_{\ts{ab}}'^{\ast} \\
\mathcal{L}_\textsc{b}' & \mathcal{L}_\textsc{bb}' & \mathcal{L}_\textsc{ab}'^\ast & 0 \\
\mathcal{L}_\textsc{a}' & \mathcal{L}_\textsc{ab}' & \mathcal{L}_\textsc{aa}' & 0 \\
\mathcal{M}_{\ts{ab}}' & 0 & 0 & 0 \\
\end{array}\right)+O(\lambda^3), \nonumber 
\end{align} 
with 
\begin{align}
    \mathcal{L}_{\ts i}' &= \frac{\epsilon}{\epsilon^2 + \mathcal{L}_{\ts{cc}}}\left(e^{-\ii \xi}\mathcal{M}_{\ts{i}\ts{c}}^* + e^{\ii \xi}\mathcal{L}_{\ts{i}\ts{c}}^*\right),  \label{eq:TrLi} \\
    \mathcal{L}_{\ts{ij}}' &= \mathcal{L}_{\ts{ij}} + \frac{\mathcal{L}_{\ts{ic}} \mathcal{L}_{\ts{jc}}^{*} + \mathcal{M}_{\ts{ic}} \mathcal{M}_{\ts{jc}}^{*}}{\epsilon^2 + \mathcal{L}_{\ts{cc}}} \;, \label{eq:TrLij} \\
    \mathcal{M}_{\ts{ab}}' &=  \mathcal{M}_{\ts{ab}} + \frac{\mathcal{L}_{\ts{ac}} \mathcal{M}_{\ts{bc}} + \mathcal{L}_{\ts{bc}} \mathcal{M}_{\ts{ac}}}{\epsilon^2 + \mathcal{L}_{\ts{cc}}} \label{eq:TrMij} \;,
\end{align}
for $\ts{I},\ts{J} \in \{\ts{A},\ts{B}\}$. Note that $\mathcal{L}_\ts{i}'=\Theta(\lambda)$, $\mathcal{L}_{\ts{ij}}'=\Theta(\lambda^2)$, and $\mathcal{M}_{\ts{ab}}'=\Theta(\lambda^2)$. Thus, in this case the matrix $\hat{\gamma}$, which yields the terms $\mathcal{L}_\textsc{i}'$, contributes to the joint state of the detectors at first order in $\lambda$. This is in contrast with both the non-orthogonal and orthogonal regimes, for which the contribution of $\hat{\gamma}$ was always of order $\lambda^{3/2}$ or higher, and we saw that it did not affect the entanglement harvested. 

Using the matrix form given in Eq.~\eqref{eq:matrixTransition}, and the methods given in Appendix~\ref{appendix negativity}, we can express the negativity of the post-selected state as
\begin{align}\label{eq:NegativityTransition}
    \mathcal{N}_s&=\max\left(\!0\,, \frac{\sqrt{\bar{\alpha}^2+\beta^2-2\bar{\beta}\bar{\alpha}+\zeta}}{2}-\frac{\alpha-\beta}{2} \right) \\
    & \phantom{=\,}+\mathcal{O}(\lambda^3) \;, \nonumber
\end{align}
where we have denoted
\begin{align}
    \alpha&=\L_{\ts{aa}}'\! +\! \L_{\ts{bb}}' \;,\\
    \bar{\alpha}&=\L_{\ts{aa}}'\! -\! \L_{\ts{bb}}' \;,\\
    \beta&=|\mathcal{L}_{\ts{a}}'|^2 + |\mathcal{L}_{\ts{b}}'|^2 \;,\\
    \bar{\beta}&=|\mathcal{L}_{\ts{a}}'|^2 - |\mathcal{L}_{\ts{b}}'|^2 \;,\\
    \zeta&=8\Re(\L_{\ts{a}}'\L_{\ts{b}}'\M_{\ts{ab}}'^*) \;.
\end{align}
We will obtain explicit results when we examine an explicit protocol in Section~\ref{explicitProtocol}.

\section{Case study: Inertial pointlike detectors with Gaussian switchings} \label{explicitProtocol}

We have seen that measurements do not affect the entanglement harvested at leading order in $\lambda$ in the non-orthogonal regime. On the other hand, for the outcomes $s$ for which $\epsilon=\braket{s}{g_\ts{c}}$ is sufficiently small, i.e., in the orthogonal and the transition regimes, the measurement has a more disruptive effect on the joint state of detectors A and B. As seen in Eqs. \eqref{eq:NegativityOrthogonal} and \eqref{eq:NegativityTransition}, in these regimes the measurement affects the entanglement harvested at leading order. In this Section we study an explicit protocol in the simplest spacetime background and detector trajectories. Namely, we will consider inertial comoving pointlike detectors in Minkowski spacetime with Gaussian switchings. 

We consider that the three particle detectors involved in the protocol are at rest in the inertial reference frame $(t,\bm x)$, and therefore their proper times coincide with the coordinate time $t$. Their trajectories can be parametrized as $\mf z_{\ts i}(t) = (t, \bm{x}_{\ts i})$,  where $\bm{x}_{\ts i}$ is constant and $\ts I \in \{\ts A, \ts B, \ts C \}$, and their energy gaps are all the same, \mbox{$\Omega_{\ts a} = \Omega_{\ts b} = \Omega_{\ts c} \equiv \Omega$}. For the pointlike case, the spacetime smearing can be written as
\begin{equation}
    \Lambda_{\ts i}(t, \bm x) = \chi_\ts{i}(t)\,\delta^{(3)}(\bm{x}-\bm{x}_\ts{i}),
\end{equation}
where $\chi_\ts{i}(t)$ is the so-called switching function, which controls the strength of the coupling of the detector and the field in time. For this case study we consider that the switching functions will be Gaussian,
\begin{align}
    \chi_{\ts i}(t) &= \frac{1}{\sqrt{2 \pi}}\exp \bigg(-\frac{(t-t_{\ts i})^2}{2 T^2}\bigg),
\end{align}
where $T$ determines the effective duration and $t_{\ts i}$ the peaks of the interaction of each detector with the field. With these choices, the interactions of the three detectors are identical in shape, and localized around different points of spacetime $(t_{\ts i}, \bm x_{\ts i})$.

Although the Gaussian switching function for detector C is non-zero for all $t \in \mathds{R}$, 
\begin{equation}
    \int_{t_\ts{c}+5T}^{+\infty}\dd t \, \chi_\ts{c}(t) < 10^{-6} \,\int_{-\infty}^{+\infty} \dd t \, \chi_\ts{c}(t) \;.
\end{equation}
Thus, if we assume that the measurement on detector C is performed at $t = t_0 \geq t_{\ts c} + 5T$, we can consider that its interaction is effectively decoupled from the field at the time at which the measurement is performed\footnote{This kind of approximation is common in relativistic quantum information and has been studied rigorously in the context of interaction regions that are effectively spacelike separated, even when the supports of the detectors' switching and smearing functions are not compact (see, e.g.,~\cite{ericksonNew,HectorTales}).}. This approximation allows us to use all the calculations\footnote{In more detail, in this case we should apply the operator $\hat{U}(+ \infty,t_0) \hat{P} \hat{U}(t_0,-\infty)$ in Eq. \eqref{eq:finalJointState} instead of $ \hat{P} \hat{U}$, where
\begin{equation}
    \hat{U}(t_2, t_1) =  \mathcal{T}\text{exp}\left(-\ii \int \dd V  \,\hat{h}(\mf x) \,\theta(t-t_1)\,\theta(t_2 - t)\right).
\end{equation}
However, if Clara's detector is effectively switched off at \mbox{$t > t_0>t_\ts{c}+4T$}, we can assume that the time evolution corresponding to detector C contained in $\hat{U}(+ \infty,t_0)$ is negligible. Since the time-evolution associated to detectors A and B commutes with $\hat{P}$ (they act on different Hilbert spaces), we have $\hat{U}(+ \infty,t_0) \hat{P} \hat{U}(t_0,-\infty) \simeq \hat{P} \hat{U}$. 
} given in Section~\ref{Setup}.

Recall that, for all regimes, the negativity is a function of the terms $\mathcal{L}_{\ts{ij}}$ and $\mathcal{M}_{\ts{ij}}$ given in Eqs.~\eqref{eq:Lij} and~\eqref{eq:Mij}. For the specific protocol presented in this Section, the $\mathcal{L}_\ts{ij}$ terms can be given in closed-form. Due to the translational symmetry of Minkowski spacetime, all the local terms are equal, $\mathcal{L}_{\ts{aa}} = \mathcal{L}_{\ts{bb}} = \mathcal{L}_{\ts{cc}} \equiv \mathcal{L}$. The single momentum integral can be solved explicitly,
\begin{align} 
    \mathcal{L} &= \frac{\lambda^2T^2}{2 \pi^2}\int_{0}^{\infty}\frac{\dd k}{2 } \, k \, e^{-(\Omega+ k)^2 T^2} \\
    &= \frac{e^{-\Omega^2 T^2} - \sqrt{\pi} \,\Omega T\, \text{erfc}(\Omega T)}{8\pi^2}. \label{req:Lvacuum}
\end{align}
 The non-local terms $\mathcal{L}_{\ts{ij}}$, for \mbox{$\ts I \neq \ts J$}, can also be cast as a single momentum integral which admits a closed form
\begin{align}\label{req:Lij integral} 
    &\mathcal{L}_{\ts{ij}} = \frac{\lambda^2T^2}{2 \pi^2}\int_{0}^{\infty}\frac{\dd k}{2 } \, k e^{-(\Omega+ k)^2 T^2 + \ii(\Omega + k)\Delta_{\ts{ij}}} \, \text{sinc}(k L_{\ts{ij}}) \nonumber \\[0.1cm]
    &= \frac{\lambda^2}{16  \pi^{3/2}}\frac{T}{L_{\ts{ij}}}e^{-\frac{ L_{\ts{ij}}^2 }{4 T} - \frac{\Delta_{\ts{ij}}^2}{T^2}}\\[0.1cm]
    &\phantom{=\,}\times\bigg\{e^{\frac{ L_{\ts{ij}} \Delta_{\ts{ij}}}{2T^2} + \ii \Omega L_{\ts{ij}}}\bigg[\ii + \operatorname{erfi}\bigg( \frac{ L_{\ts{ij}}- \Delta_{\ts{ij}}}{T^2} - \ii \Omega T\bigg)\bigg] \nonumber\\[0.1cm]
    &\phantom{=\,} + e^{\frac{- L_{\ts{ij}} \Delta_{\ts{ij}}}{2 T^2} - \ii \Omega L_{\ts{ij}}}\bigg[-\ii + \operatorname{erfi}\bigg( \frac{ L_{\ts{ij}}+ \Delta_{\ts{ij}}}{T^2} + \ii \Omega T\bigg) \bigg]\bigg\}, \nonumber 
\end{align}
where $L_\ts{ij}=|\bm x_\ts{i}-\bm x_\ts{j}|$ is the distance between detectors $\ts I$ and $\ts J$, and $\Delta_\ts{ij}=t_\ts{i}-t_\ts{j}$ is the delay between the peaks of their switching functions. The $\mathcal{M}_{\ts{ij}}$ terms can also be written as a momentum integral, but they do not admit a closed form:
\begin{align}\label{req:Mvacuum}
    \mathcal{M}_\ts{ij} &= -\frac{\lambda^2T^2}{4 \pi^2}e^{\ii \Omega(t_\ts{i} + t_\ts{j})}\int_{0}^{+\infty}\frac{\dd k}{2 } \, k \, e^{-(\Omega^2+ k^2) T^2} \text{sinc}(k L_{\ts{ij}})\nonumber \\
    &\phantom{=\,}\times \bigg\{e^{\ii \Omega \Delta_{\ts{ij}}}\bigg[1 - \operatorname{erf}\bigg( \frac{ \Delta_{\ts{ij}}}{2 T} + \ii k\bigg) \bigg] \\
    &\phantom{====}+ e^{-\ii \Omega \Delta_{\ts{ij}}}\bigg[1 - \operatorname{erf}\bigg( \frac{ -\Delta_{\ts{ij}}}{2T} + \ii k\bigg) \bigg]\bigg\}. \nonumber 
\end{align}
 Nevertheless, these terms can be easily evaluated numerically, which will allow us to analyze the effect of the measurement in the orthogonal and the transition regimes.  

\subsection{Orthogonal regime}

In this Subsection, we consider several entanglement harvesting scenarios, and we study the effect that the measurement has on the entanglement harvested when we are in the orthogonal regime. Thus, in these scenarios the negativity is given by Eq.~\eqref{eq:NegativityOrthogonal}. 

\begin{figure}[h]
    \includegraphics[scale=0.75]{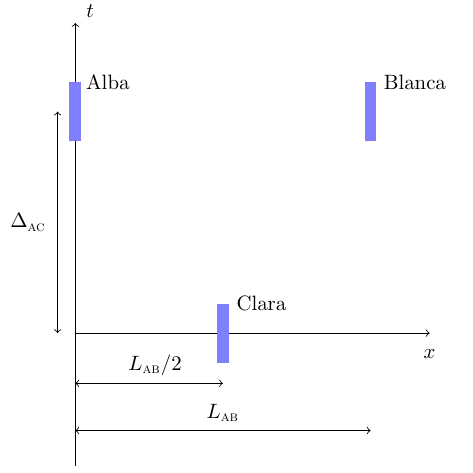}
    \caption{Configuration in a slice of spacetime of the interaction regions (in blue) of the three detectors involved in the first example protocol, i.e., with parameters \mbox{$L_{\ts{ac}} = L_{\ts{bc}} = L_{\ts{ab}}/2 = 2.5T$} and \mbox{$\Delta_{\ts{ab}} = 0$}.}\label{fig: scheme1}
\end{figure}

First, we consider a protocol in which Alba and Blanca harvest spacelike entanglement from the quantum field. In this setup, their detectors are switched on at the same coordinate time, that is $\Delta_{\ts{ab}} = 0$, and we take the proper distance between them to be $L_{\ts{ab}}$. We take detector C to be located at the middle point between detectors A and B, setting $L_{\ts{ac}} = L_{\ts{bc}} = L_{\ts{ab}}/2$ (see Fig.~\ref{fig: scheme1}). 

\begin{figure}[h]
    \includegraphics[scale=0.73]{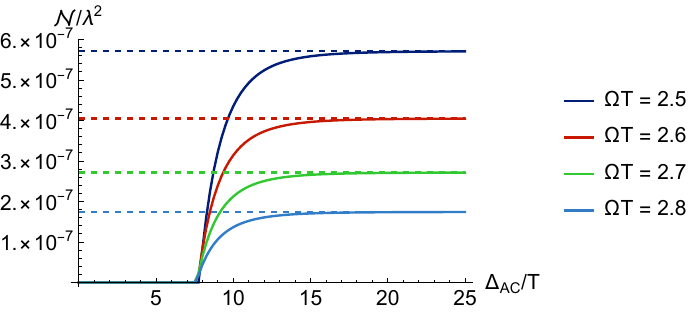}
    \caption{(Solid lines) Negativity in the orthogonal regime of $\rhoh_{\ts{ab}}^{s}$ as a function of the delay $\Delta_{\ts{ac}}$ for different values of the energy gap $\Omega T$, and parameters \mbox{$L_{\ts{ac}} = L_{\ts{bc}} = L_{\ts{ab}}/2 = 2.5T$} and \mbox{$\Delta_{\ts{ab}} = 0$}. This plot corresponds to the scheme in Fig. \ref{fig: scheme1}.
    (Dashed lines) Negativity of $\rhoh_{\ts{ab}}$ for the same values of $L_{\ts{ab}}$ and $\Delta_{\ts{ab}}$ for different $\Omega T$.}\label{fig:spacelikeO}
\end{figure}

If we fix $L_{\ts{ab}} = 5T$, the interaction regions of the detectors $\ts{A}$ and $\ts{B}$ can be considered to be effectively spacelike separated for entanglement harvesting purposes. Namely, it was shown in \cite{HectorTales} that for this specific setup, communication between the detectors has a negligible effect in the entanglement that they harvest (see also \cite{ericksonNew}). The negativity of $\rhoh_{\ts{ab}}^s$ in this configuration is plotted in Fig.~\ref{fig:spacelikeO} as a function of the delay $\Delta_{\ts{ac}} = \Delta_{\ts{bc}}$, for different energy gaps. We consider positive values for the delay, so that this configuration can be understood, e.g., as a simplified model for a spacelike entanglement harvesting protocol in which a  measurement is performed during the preparation of the experiment. For comparison, we also plotted with dashed lines the negativity of $\rhoh_\ts{ab}$, which corresponds to the entanglement harvested with the same protocol for Alba and Blanca, but without Clara's intervention, i.e., without measurements. Several values of $\Omega$ were chosen to see that the effect of the measurement does not qualitatively change as a function of the energy gap. As a baseline for comparison, we used the value of $\Omega$ for which the maximum of the negativity of $\rhoh_{\ts{ab}}$ is reached, which was found in~\cite{HectorTales} to be approximately $\Omega_{\text{max}}\approx L_{\ts{ab}}/2T^2 = 2.5/T$. For each value of $\Omega T$, the negativity of $\rhoh_{\ts{ab}}^s$ is always below the negativity of $\rhoh_{\ts{ab}}$. Therefore, the measurement decreases the entanglement harvested. We observe that if the delay between the switching function for C and the switching functions for A and B is small enough, the measurement destroys the entanglement acquired by A and B up to second order in $\lambda$ for every examined case. The negativity of $\rhoh_{\ts{ab}}^s$ increases with $\Delta_{\ts{ac}}$. For large values of $\Delta_{\ts{ac}}$, the negativity of $\rhoh_{\ts{ab}}^s$ tends asymptotically to the negativity of $\rhoh_{\ts{ab}}$. This is because the field correlations of the interaction region of C with the interaction regions of A and B decay as the delay between them increases. Indeed, as $\Delta_{\ts{ac}} = \Delta_{\ts{bc}}$ increases, the non-local terms $\mathcal{L}_\ts{ic}$ and $\mathcal{M}_\ts{ic}$ involving detector C decrease, and we have that $\Tilde{\L}_{\ts{ij}}$ and $\Tilde{\M}_{\ts{ij}}$ tend to $\L_{\ts{ij}}$ and $\M_{\ts{ij}}$, respectively (see Eqs.~\eqref{eq:TLij} and~\eqref{eq:TMij}). Therefore, the effect of the measurement on the entanglement harvested becomes negligible for large enough delays. 

The reader might be surprised by the fact that for configurations in which A, B and C are supposed to be spacelike separated, the measurement performed by Clara on detector C can end up not only affecting the entanglement harvested by A and B, but completely destroying it. One can make sense of this taking into account that the state $\rhoh_\ts{ab}^{s}$ only has physical meaning in a region of spacetime when the correlations between A and B can be measured. This necessarily corresponds to a ``processing region''~\cite{Ruep2021,Jose} in the causal future of both Alba and Blanca, which, given the configuration in Fig.~\ref{fig: scheme1}, will also be in the causal future of Clara's measurement. The outcome of the measurement will therefore be available to the processing region, and since detector C is correlated with the field, which is correlated with detectors A and B, the corresponding update does indeed affect the joint state of A and B, even if both the interaction and the measurement of C are spacelike separated from the interactions of detectors A and B with the field. 

\begin{figure}[h]
    \includegraphics[scale=0.73]{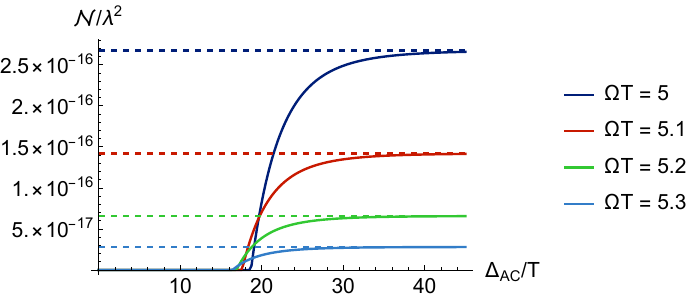}
    \caption{(Solid lines) Negativity in the orthogonal regime of $\rhoh_{\ts{ab}}^s$ as a function of the delay $\Delta_{\ts{ac}}$ for different values of the energy gap $\Omega T$, and parameters $L_{\ts{ac}} = L_{\ts{bc}} = L_{\ts{ab}}/2 = 5T$ and $\Delta_{\ts{ab}} = 0$. This plot corresponds to the scheme in Fig. \ref{fig: scheme1}. (Dashed lines) Negativity of $\rhoh_{\ts{ab}}$ for the same values of $L_{\ts{ab}}$ and $\Delta_{\ts{ab}}$ for different $\Omega T$.} \label{fig:spacelikeOLejos}
\end{figure}

Note that with the choice of parameters analyzed in Fig.~\ref{fig:spacelikeO}, detectors A and B interact with the field in effectively spacelike separated regions. However, due to its position, C is in effective causal contact with both A and B. Because of this, in this case one might wonder then if the leading order effect of the measurement is due to the interaction of the detector with the field (which then propagates to the effective interaction regions of detectors A and B), rather than to the projective measurement performed on the detector and the field correlations. In order to clarify this, we analyzed the behaviour of the negativity for the same configuration depicted in Fig.~\ref{fig: scheme1}, but with a larger separation between the particle detectors: \mbox{$L_\ts{ac}=L_\ts{bc}=L_\ts{ab}/2=5T$}. The results are plotted in Fig.~\ref{fig:spacelikeOLejos}. For this choice of parameters, we can consider that as $\Delta_\ts{ac}=\Delta_\ts{bc}$ approaches zero, the interaction region of detector C is effectively spacelike separated from those of detectors A and B. We then observe that even when the three interactions are effectively causally disconnected, the measurement does destroy the entanglement harvested. We conclude that the three detectors get correlated due to their interactions with the field, and it is indeed Clara's measurement what sabotages the entanglement between A and B. 

Notice that the fact that the selective measurement on C affects A and B does not mean that it can be used to send information from C to A and B. Indeed, the reason the measurement on C affects the joint state of the detectors A and B is that we are post-selecting the state to having obtained a particular outcome of the measurement of C. Thus, the effect we are witnessing when a spacelike measurement on C affects A and B is not one of faster-than-light signalling, but one of correlations between the state of C (and therefore the outcome of the measurement performed on it) and the joint state of A and B.

\begin{figure}[]
    \includegraphics[scale=0.78]{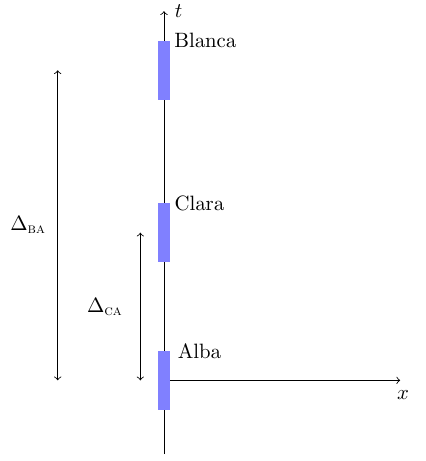}
    \caption{Configuration in a slice of spacetime of the interaction regions (in blue) of the three detectors involved in the first example protocol, i.e., with parameters \mbox{$L_{\ts{ac}} = L_{\ts{bc}} = 0$}.}\label{fig: scheme2}
\end{figure}

We now analyze the effect of the measurement in a timelike entanglement harvesting protocol. Notice that due to the strong Huygens principle in 3+1 spacetime dimensions, when the detectors are in strictly timelike separation, the entanglement acquired between is not due to communication but rather is genuine harvesting of the field correlations (see, e.g.~\cite{ericksonNew}). For simplicity, we set $L_{\ts{ab}} = L_{\ts{ac}} = L_{\ts{bc}} = 0$, and we keep constant the delay between the switching functions of Alba and Blanca to be $\Delta_{\ts{ab}}$. For this setup, we vary the delay $\Delta_{\ts{ac}} > 0$, as depicted in Fig.~\ref{fig: scheme2}.

\begin{figure}[h]
    \includegraphics[scale=0.73]{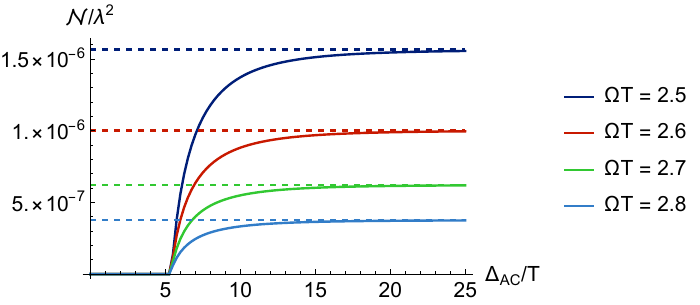} 
    \caption{(Solid lines) Negativity in the orthogonal regime of $\rhoh_{\ts{ab}}^s$ as a function of the delay $\Delta_{\ts{ac}} > 0$ for different values of the energy gap $\Omega T$, and parameters $L_{\ts{ac}} = L_{\ts{bc}} = L_{\ts{ab}} = 0$ and  $\Delta_{\ts{ab}} = 5T$. This plot corresponds to the scheme in Fig. \ref{fig: scheme2}. (Dashed lines) Negativity of $\rhoh_{\ts{ab}}$ for the same values of $L_{\ts{ab}}$ and $\Delta_{\ts{ab}}$ for different $\Omega T$.}\label{fig:timelikeO}
\end{figure}

The negativity of $\rhoh_{\ts{ab}}^s$ in the orthogonal regime for this configuration is plotted in Fig.~\ref{fig:timelikeO} as a function of the delay $\Delta_{\ts{ac}} > 0$, for different values of the energy gap. This configuration could model, e.g., a timelike entanglement harvesting protocol for which a measurement has been performed during its preparation. We also plot the negativity of $\rhoh_{\ts{ab}}$ corresponding to the protocol when Clara does not carry out any measurement. We observe a similar behaviour to the spacelike entanglement harvesting protocol previously analyzed. In particular, the measurement destroys the entanglement harvested at second order by Alba and Blanca for small enough $\Delta_{\ts{ac}}$. As $\Delta_{\ts{ac}}$ increases, the detectors start harvesting entanglement, and the negativity of $\rhoh_{\ts{ab}}^s$ tends to that of $\rhoh_{\ts{ab}}$ for large values of $\Delta_{\ts{ac}}$, for the same reason as in the case of the spacelike harvesting configuration in Fig.~\ref{fig:spacelikeO}.

\begin{figure}[h]
    \includegraphics[scale=0.78]{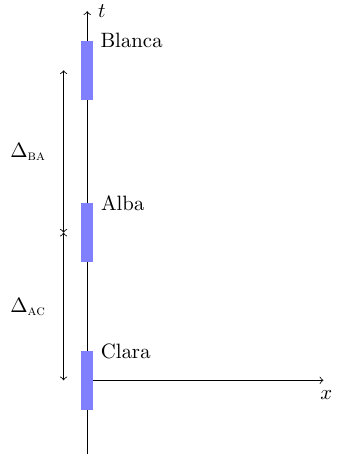}
    \caption{Configuration in a slice of spacetime of the interaction regions (in blue) of the three detectors involved in the first example protocol, i.e., with parameters \mbox{$L_{\ts{ac}} = L_{\ts{bc}} = L_{\ts{ab}} = 0$}. }\label{fig: scheme3}
\end{figure}


Finally, we also analyze the case in which detector C interacts with the field at a time between the interactions of A and B, as shown in Fig. \ref{fig: scheme3}. We fixed $\Delta_{\ts{ba}} = 5T$. For all the values of $\Delta_{\ts{ac}}$ and for all the different energy gaps studied, we find that the measurement cancels the entanglement between detectors A and B up to second order. 

Overall, in all the configurations studied we have associated the measurement in the orthogonal regime with a decrease or even a complete cancellation of the entanglement harvested. However, if the delay between the switching function of detector C and the switching functions of the rest of the detectors is large enough, the effect of the measurement is suppressed. This is because the field correlations between the interaction region of C and the region where harvesting is performed decay with the spacetime separation between them, so that the selective measurement reveals increasingly less information about the field at the harvesting region.

\subsection{Transition regime}

In this Subsection, we analyze the effect that the measurement has on the entanglement harvested when we are in the transition regime, for the same configurations studied in the previous Subsection. Thus, in these scenarios negativity is given by Eq.~\eqref{eq:NegativityTransition}.

As discussed in Section~\ref{regimes}, the matrix $\hat{\gamma}$ only yields a leading order contribution to the negativity in the transition regime. We will see that this contribution yields a more complex behaviour than both the non-orthogonal regime, in which the negativity is not affected at leading order, and the orthogonal regime, in which the measurement decreases the negativity. In particular, in this case the entanglement harvested can be increased by a third party performing field measurements.

Recall that in the transition regime we analyzed the case $\epsilon = \Theta(\lambda)$, and note that in this case the specific value of $\epsilon$ affects $\hat\rho_\ts{ab}^s$ at leading order (see Eqs.~\eqref{eq:matrixTransition}--\eqref{eq:TrMij}). For simplicity, in what follows we take $\epsilon=\lambda$.

\begin{figure}[h]
    \includegraphics[scale=0.73]{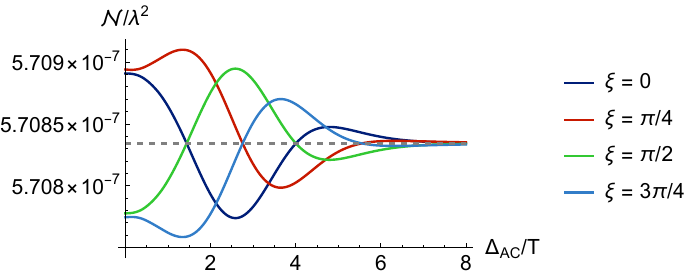} 
    \caption{(Solid lines) Negativity in the transition regime of $\rhoh_{\ts{ab}}^s$ as a function of  $\Delta_{\ts{ac}}$ for different values of $\xi$, and \mbox{$\Omega T=2.5$, $L_{\ts{ac}} = L_{\ts{bc}} = L_{\ts{ab}}/2 = 2.5T$} and $\Delta_{\ts{ab}} = 0$. This plot corresponds to the scheme in Fig. \ref{fig: scheme1}. (Dashed line) Negativity of $\rhoh_{\ts{ab}}$ for the same values of $L_{\ts{ab}}$, $\Delta_{\ts{ab}}$ and $\Omega T$.}\label{fig:spacelikeT}
\end{figure}

We first analyze the contribution of the measurement in a scenario of spacelike entanglement harvesting (the situation on Fig.~\ref{fig: scheme1}). As before, we set $\Delta_{\ts{ab}} = 0$ and \mbox{$L_{\ts{ac}} = L_{\ts{bc}} = L_{\ts{ab}}/2 = 2.5 T$}. The negativity for this configuration is plotted in Fig.~\ref{fig:spacelikeT} as a function of the delay between the switching function of detectors A and C, $\Delta_{\ts{ac}} > 0$, for different values of the phase $\xi$ (see Eq.~\eqref{eq:outcome s}) and fixed energy gap $\Omega T = 2.5$. We also plot the negativity of $\rhoh_{\ts{ab}}$ for the same configuration. The measurement introduces small damped oscillations with the parameter $\Delta_{\ts{ac}}$ that converge for large $\Delta_{\ts{ac}}$ to the negativity of $\rhoh_{\ts{ab}}$, with the phase of the oscillations controlled by $\xi$. It is worth pointing out that the relative difference between the negativity of $\rhoh_{\ts{ab}}^s$ and the negativity of $\rhoh_{\ts{ab}}$ is never larger than a $0.1\%$. This may however be related to the fact that we are working in a weak measurements regime, and therefore using a stronger coupling between detector C and the field might yield a higher relative enhancement.

In any case, we see that the effect is small. In fact, we performed the analysis for a different set of parameters, namely, $\Delta_{\ts{ab}} = 0$ and $L_{\ts{ac}} = L_{\ts{bc}} = L_{\ts{ab}}/2 = 5T$, with a fixed energy gap of $L_{\ts{ab}}/2T^2 = 5/T$, observing that in this case it is not possible to distinguish numerically the negativity of $\rhoh_{\ts{ab}}$ and the negativity of $\rhoh_{\ts{ab}}^s$. This is related to the fact that the terms $\L_{\ts{ij}}$ and $\M_{\ts{ij}}$ decay with the energy gap. As seen in Eqs.~\eqref{eq:TrLi}--\eqref{eq:NegativityTransition}, the measurement contributions to the negativity are given by products of the form $\mathcal{L}_\ts{ic}\mathcal{L}_\ts{jc}^*$, $\mathcal{M}_\ts{ic}\mathcal{M}_\ts{jc}^*$, and $\mathcal{L}_\ts{ic}\mathcal{M}_\ts{jc}$. Meanwhile, the contributions due to the entanglement harvesting protocol performed by A and B, which are present both with and without measurements, are given in terms of $\mathcal{L}_\ts{ij}$ and $\mathcal{M}_\ts{ab}$. As a result, the measurement contributions decay faster with the energy gap. Since we have considered a value of the gap that is double\footnote{This is because for larger separation distances one needs to consider larger energy gaps to harvest entanglement~\cite{Pozas-Kerstjens:2015,Pozas2016}.} the value used for the analysis in Fig.~\ref{fig:spacelikeT}, the effect of the measurement is suppressed. This is in contrast with the behaviour displayed in the orthogonal regime, when the energy gap does not modify the relative magnitude of the effect of the measurement in the negativity of $\rhoh_{\ts{ab}}^s$ with respect to $\rhoh_\ts{ab}$. 
 
\begin{figure}[h]
    \includegraphics[scale=0.73]{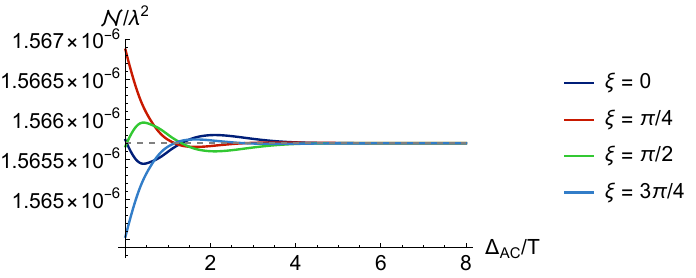}
    \caption{(Solid lines) Negativity in the transition regime of $\rhoh_{\ts{ab}}^s$ as a function of  $\Delta_{\ts{ac}}$ for different values of  $\xi$, and  $\Omega T=2.5$, $L_{\ts{ac}} = L_{\ts{bc}} = L_{\ts{ab}} = 0$ and $\Delta_{\ts{ba}} = 5T$. This plot corresponds to the scheme in Fig.~\ref{fig: scheme2}. (Dashed line) Negativity of $\rhoh_{\ts{ab}}$ for the same values of $L_{\ts{ab}}$, $\Delta_{\ts{ba}}$ and $\Omega T$.} \label{fig:timelikeT}
\end{figure}

We now analyze the effect of the measurement in a timelike  harvesting protocol (see Fig.~\ref{fig: scheme2}). As for the orthogonal regime, we set \mbox{$L_{\ts{ab}} = L_{\ts{ac}} = L_{\ts{bc}} = 0$} and \mbox{$\Delta_{\ts{ba}} = 5T$}. The negativity of $\rhoh_{\ts{ab}}$ is plotted in Fig.~\ref{fig:timelikeT} as a function of $\Delta_{\ts{ac}} >0$ for different values of $\xi$ and for $\Omega T = 2.5$. As for the spacelike protocol, we observe damped oscillations that converge to the negativity of $\rhoh_{\ts{ab}}$ for large values of the delay $\Delta_{\ts{ac}}$. The phase of the oscillations is determined by the parameter $\xi$. 

\begin{figure}[h]
    \includegraphics[scale=0.73]{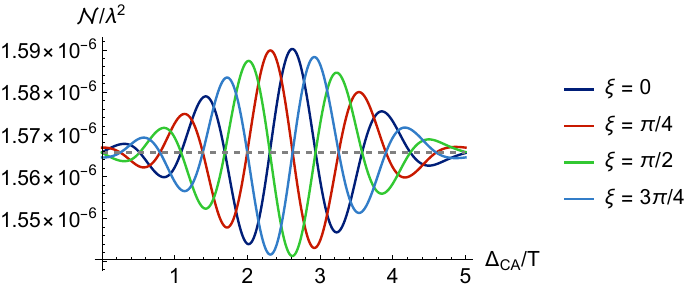}
    \caption{(Solid lines) Negativity in the transition regime of $\rhoh_{\ts{ab}}^s$ as a function of the delay $0<\Delta_{\ts{ca}}<\Delta_\ts{ba}$ for different values of the phase $\xi$, and parameters $\Omega T=2.5$, \mbox{$L_{\ts{ac}} = L_{\ts{bc}} = L_{\ts{ab}} = 0$} and $\Delta_{\ts{ba}} = 5T$. This plot corresponds to the scheme in Fig. \ref{fig: scheme2}. (Dashed line) Negativity of $\rhoh_{\ts{ab}}$ for the same values of $L_{\ts{ab}}$, $\Delta_{\ts{ba}}$ and $\Omega T$.} \label{fig:timelike2T}
\end{figure}

Finally, Fig.~\ref{fig:timelike2T} shows the analysis for a timelike protocol with \mbox{$0 < \Delta_{\ts{ca}} < \Delta_{\ts{ba}} = 5T$} (see scheme  in Fig. \ref{fig: scheme3}). A similar oscillatory  behaviour of the negativity with $\Delta_{\ts{ca}}$ is observed. The amplitude of the oscillations is maximized if the delays $\Delta_{\ts{ca}}$ and $\Delta_{\ts{bc}}$ are equal. 

Overall, we have associated the effect of the measurement in the transition regime with small modifications of the negativity at leading order, which are suppressed with the energy gap. The measurement introduces oscillations with the delay between the switching function of detector C and the switching functions of detectors A and B.

\section{Non-selective measurement} \label{Non-selective}
In previous sections we have discussed protocols in which the measurement taking place is a selective one, i.e., the outcome is known and the state of the system is updated accordingly. However, it is also possible to consider the case in which the measurement performed by Clara is non-selective. The update in this case only takes into account that the measurement is performed, but not the outcome, since this information is assumed to be unavailable. This update is the weighted average of the updates associated with all the possible outcomes, where the weights are their corresponding probabilities of occurrence. Let us denote with $\ket{s}, \ket{\bar{s}}$ the eigenstates associated with the possible outcomes of the measurement, which form an orthonormal basis of the Hilbert space of detector C. The non-selectively updated joint detectors-field state $\rhoh^{\ts{NS}}$ is then given by
\begin{align}
    \rhoh^{\text{NS}} = \textrm{Prob}(s)\,\rhoh^{s} + \textrm{Prob}(\bar{s})\,\rhoh^{\bar{s}},
\end{align}
where $\rhoh^s$ and $\rhoh^{\bar{s}}$ are the updates corresponding to each outcome, given by Eq.~\eqref{eq:selective}, and $\textrm{Prob}(s)$ and $\textrm{Prob}(\bar{s})$ are their probabilities, given by Eq.~\eqref{eq:denominator}. This update corresponds to an observer who is aware of the measurement being performed but who is not aware of its outcome~\cite{Luders1951,Hellwig1970formal,Jose}. Alba and Blanca would use this update if Clara is spacelike separated from them but she told them beforehand that she would perform the measurement. 
The above equation can be expanded as
\begin{align}
    \rhoh^{\text{NS}} = \proj{s}{s}\!\hat{U} \rhoh^{(0)}\hat{U}^{\dagger} \!\proj{s}{s} + \proj{\bar{s}}{\bar{s}}\!\hat{U} \rhoh^{(0)}\hat{U}^{\dagger}\! \proj{\bar{s}}{\bar{s}}.
\end{align}
Tracing out the degrees of freedom corresponding to the field and detector C, we recover the partial state of detectors A and B,
\begin{align}
    \rhoh_{\ts{ab}}^{\text{NS}} &= \text{Tr}_{\ts{c},\phi}\left( \rhoh^{\text{NS}} \right) \nonumber\\
    &= \bra{s}\!\hat{U} \rhoh^{(0)}\hat{U}^{\dagger} \!\ket{s} + \bra{\bar{s}}\!\hat{U} \rhoh^{(0)}\hat{U}^{\dagger}\! \ket{\bar{s}}
    \\
    &= \text{Tr}_{\ts{c},\phi}\Big(\hat{U} \rhoh^{(0)}\hat{U}^{\dagger} \Big). \nonumber
\end{align}
 This is in fact the same final partial state that would be obtained if the three detectors had interacted with the field and the non-selective projective measurement on detector C had not been performed at all~\cite{Jose}. Moreover, this state does not contain any term related to detector C up to second order in $\lambda$. Concretely,
 \begin{equation}
     \rhoh_{\ts{ab}}^{\text{NS}} =  \rhoh_{\ts{ab}}^{(0)} + \rhoh_{\ts{ab}}^{(2)} + \mathcal{O}(\lambda^4),
 \end{equation}
where $\rhoh_{\ts{ab}}^{(2)}$ is given in Eq.~\eqref{eq:rho2}. This is, up to second order in $\lambda$, the same joint state for detectors A and B that we get when there is no detector C involved in the protocol, i.e., $\rhoh_\ts{ab}$ given in Eq.~\eqref{eq:state wo Clara}. Therefore, the negativity for the non-selective measurement at leading order is given by Eq. \eqref{eq:negativity0}. The entanglement harvested in the non-selective measurement is not affected by neither the interaction of detector C with the field nor the (non-selective) projective measurement performed on it.

\vspace{-2.5mm}

\section{Conclusions} \label{Conclusions}


In this paper, we have explored how entanglement harvesting is affected by the performance of selective and non-selective measurements of the quantum field---as modelled in~\cite{Jose}, i.e., using particle detectors. For weakly coupled detectors with identical coupling strengths, we found that the effect of the measurement at leading order is strongly dependent on the relation between the coupling parameter of the detector used to measure the field and the outcome of the measurement. Specifically, we identified three different regimes depending on the relation between the coupling strength $\lambda$ and the modulus of the scalar product between the post-measurement state on the initial state of the probe used to measure the field $\epsilon$, which is a measure of how (non-)orthogonal these two states are to each other.

For $\epsilon \gg \lambda$ (\textit{non-orthogonal regime}), we found that the measurement does not affect the entanglement harvested at leading order---even though it does affect the joint state of the particle detectors harvesting entanglement up to second order in $\lambda$, which is the leading order of the negativity. For $\epsilon \ll \lambda$ (\textit{orthogonal regime}), we studied numerically specific setups, finding that the measurement acts as a source of noise that sabotages the entanglement harvested. For the same setups, we saw that if $\epsilon \sim \lambda$ (\textit{transition regime}), the measurement has a small effect that can in principle yield small enhancements of the entanglement harvested for an adequate choice of parameters of the protocol. We noted that the effects of the measurement are suppressed as the time interval between the measurement and the entanglement harvesting protocol increases. Finally, we studied the effect of non-selective measurements, to find that they do not affect the entanglement harvested at leading order. This makes clear that the disruptive effects of the measurement on the entanglement harvested observed in the orthogonal and transition regimes are due to post-selection.

Since our study reveals that the measurement protocol affects the results of entanglement harvesting, it is arguable that the process of state preparation and measurement may be important in an experimental implementation of the protocol. As such, analyses of the kind performed  in this paper may be relevant for future experimental explorations of this relativistic quantum information protocol.
\vspace{-4mm}

\section{Acknowledgements}

HMG has been partially
supported by the mobility grants program of Centre de Formació
Interdisciplinària Superior (CFIS)--Universitat Politècnica de
Catalunya (UPC). JPG is supported by a Mike and Ophelia Lazaridis Fellowship. JPG also received the support of a fellowship from ``La Caixa'' Foundation (ID 100010434, with fellowship code LCF/BQ/AA20/11820043). EMM acknowledges support through the Discovery Grant Program of the Natural Sciences and Engineering Research Council of Canada (NSERC). EMM also acknowledges support of his Ontario Early Researcher award. Research at Perimeter Institute is supported in part by the Government of Canada through the Department of Innovation, Science and Industry Canada and by the Province of Ontario through the Ministry of Colleges and Universities.

\onecolumngrid
\appendix
\section{Measurement updated density operator}\label{appendix numerator} 
 
In this Appendix we prove that $\text{Tr}_{\phi,\ts{c}}\big[\hat{P}\, \hat{U}\, \rhoh^{(0)}\, \hat{U}^{\dagger}\, \hat{P}\,\big]$ admits the expansion used in Eq.~\eqref{eq:numerator}. First, we note that
\begin{align}
    \text{Tr}_{\phi,\ts{c}}\big[\hat{P}\, \hat{U}\, \rhoh^{(0)}\, \hat{U}^{\dagger}\, \hat{P}\,\big] &=  \bra{s}\, \text{Tr}_{\phi}\big[\,\hat{U}\, \rhoh^{(0)}\, \hat{U}^{\dagger}\,\big] \ket{s}= \bra{s} \rhoh_{\ts{abc}} \ket{s},
\end{align}
where $\rhoh_{\ts{abc}} = \text{Tr}_{\phi}\big[\,\hat{U}\, \rhoh^{(0)}\, \hat{U}^{\dagger}\,\big]$. Using the Dyson expansion of the time-evolution operator given in Eq.~\eqref{eq:Dyson}, $\rhoh_{\ts{abc}}$ can be written as
\begin{equation}
    \rhoh_{\ts{abc}} = \rhoh_{\ts{abc}}^{(0)} + \rhoh_{\ts{abc}}^{(2)} + \rhoh_{\ts{abc}}^{(4)} + \mathcal{O}(\lambda^6),
\end{equation}
where $\rhoh_{\ts{abc}}^{(k)}$ is proportional to $\lambda^k$:
\begin{align}
     \rhoh_{\ts{abc}}^{(2)} &= \text{Tr}_{\phi}\big[\,\hat{U}^{(2)}\rhoh^{(0)} \big] + \text{Tr}_{\phi}\big[\,\hat{U}^{(1)}\rhoh^{(0)} \hat{U}^{(1)\dagger}\,\big] + \text{Tr}_{\phi}\big[\,\rhoh^{(0)} \hat{U}^{(2)\dagger}\,\big], \\
     \rhoh_{\ts{abc}}^{(4)} &=  \sum_{i = 0}^4 \text{Tr}_{\phi}\big[\,\hat{U}^{(i)}\rhoh^{(0)} \hat{U}^{(4-i)\dagger}\,\big]. 
\end{align}
Notice that the odd-order corrections cancel because all the odd-point functions of the vacuum are zero. Using that detector C is initially in its ground state, and Eq. \eqref{eq:outcome s} for the outcome state $\ket{s}$, we have that
\begin{equation} \label{eq:zeroth}
    \bra{s} \rhoh_{\ts{abc}}^{(0)} \ket{s} = \epsilon^2 \rhoh^{(0)}_{\ts{ab}}.
\end{equation}
Regarding the second order terms, we have that
\begin{align}
    \bra{s} \rhoh_{\ts{abc}}^{(2)} \ket{s} &= \sum_{\ts{I,J} \in \{\ts{A},\ts{B}, \ts{C} \}}\lambda^2 \bigg(\int \dd V \dd V' \,\Lambda_{\ts i}(\mf x) \Lambda_{\ts j}(\mf x')\, \bra{s}\muh_{\ts j}(\tau_{\ts j}') \rhoh_{\ts{abc}}^{(0)} \muh_{\ts i}(\tau_{\ts i})\ket{s}\, W(\mf x, \mf x')  \nonumber \\  
    &\phantom{==========}-\int \dd V \dd V' \,\Lambda_{\ts i}(\mf x) \Lambda_{\ts j}(\mf x')\, \bra{s}\muh_{\ts i}(\tau_{\ts i}) \muh_\ts{j}(\tau_\ts{j}')  \rhoh_{\ts{abc}}^{(0)}\ket{s} \,W(\mf x, \mf x') \,\theta(t-t')  \nonumber \\[2mm]  
    &\phantom{==========}-\int \dd V \dd V' \,\Lambda_{\ts i}(\mf x) \Lambda_{\ts j}(\mf x')\,  \bra{s}\rhoh_{\ts{abc}}^{(0)}  \muh_{\ts j}(\tau_{\ts j}') \muh_{\ts i}(\tau_{\ts i}) \ket{s}\,W(\mf x', \mf x)\, \theta(t-t') \,\bigg). 
\end{align}
In the expression above, we analyze separately the terms which only involve the detectors harvesting entanglement, $\ts{A}$ and $\ts{B}$, the local terms associated to the detector used to perform the measurement, $\ts{C}$, and the cross-terms involving the pairs $(\ts{A},\ts{C})$ and $(\ts{B},\ts{C})$. Namely, we use 
\begin{equation} \label{eq:sumas}
    \sum_{\ts{I,J} \in \{\ts{A},\ts{B}, \ts{C} \}} = \sum_{\ts{I,J} \in \{\ts{A},\ts{B} \}}+\sum_{\ts{I,J} \in \{ \ts{C} \}} + \sum_{\substack{\ts{I,J} \in \{\ts{A}, \ts{C} \}\\ \ts{I} \neq \ts{J}}} 
    + \sum_{\substack{\ts{I,J} \in \{\ts{B}, \ts{C} \}\\ \ts{I} \neq \ts{J}}}.
\end{equation}
The terms involving detectors $\ts{A}$ and $\ts{B}$ yield  $\epsilon^2 \rhoh_{\ts{ab}}^{(2)}$, with $\rhoh_{\ts{ab}}^{(2)}$ given by Eq.~\eqref{eq:rho2}. The local terms in $\ts{C}$ yield \mbox{$ \mathcal{L}_{\ts{cc}}(1-2\epsilon^2) \rhoh^{(0)}_{\ts{ab}}$}. The last two summands in the expression above yield the correlation matrix $\hat{\gamma}$ given by Eq.~\eqref{eq:gamma}. $\bra{s} \rhoh_{\ts{abc}}^{(2)} \ket{s}$ is thus obtained adding all the contributions:
\begin{align} \label{eq:second}
    \bra{s} \rhoh_{\ts{abc}}^{(2)} \ket{s} = \epsilon^2 \rhoh_{\ts{ab}}^{(2)} + \mathcal{L}_{\ts{cc}}(1-2\epsilon^2)\rhoh^{(0)}_{\ts{ab}} + \hat{\gamma}\;.
\end{align}
It is left to calculate 
 \begin{equation}  \label{eq:A14}
      \bra{s}\rhoh_{\ts{abc}}^{(4)} \ket{s} =  \sum_{i = 0}^4 \bra{s}\text{Tr}_{\phi}\big[\,\hat{U}^{(i)}\rhoh^{(0)} \hat{U}^{(4-i)\dagger}\,\big]\ket{s}.
 \end{equation}
In order to compute the negativity at leading order in perturbation theory for the cases in which $\epsilon$ is sufficiently small (i.e., in the orthogonal and transition regimes), we only need to calculate up to fourth order terms in $\lambda$. In particular, in these regimes the only fourth order terms that influence the negativity are those which are not proportional to any power of $\epsilon$.
Since \mbox{$|\!\braket{s}{g_{\ts{c}}}\!| = \epsilon$}, we have that \mbox{$|\!\braket{s}{e_{\ts c}}\!| = \sqrt{1-\epsilon^2} = 1 + \mathcal{O}(\epsilon^2)$}, and thus we only need to compute the terms for which the time-evolution yields $\ketbra{e_{\ts c}}{e_{\ts c}}$. These terms are characterized by having the Hamiltonian weight $\hat{h}_{\ts c}$ acting an odd number of times on each side of the initial state $\rhoh_{\ts{abc}}^{(0)}$. Thus, we need not to calculate \mbox{$\bra{s}\text{Tr}_{\phi}\big[\,\hat{U}^{(0)}\rhoh^{(0)} \hat{U}^{(4)\dagger}\,\big]\ket{s}$} and \mbox{$\bra{s}\text{Tr}_{\phi}\big[\,\hat{U}^{(4)}\rhoh^{(0)} \hat{U}^{(0)\dagger}\,\big]\ket{s}$} since they only yield terms proportional to $\epsilon$. The only terms that yield contributions of order zero in $\epsilon$ are those with a single Hamiltonian weight $\hat{h}_{\ts{c}}(\mf x)$ acting on each side of $\rhoh_{\ts{abc}}$: 
\begin{align}\label{appA}
    \bra{s}\hat{U}^{(2)}\! \,\rhoh_{\ts{abc}}^{(0)}\, \hat{U}^{(2)\dagger}\ket{s}&=\bra{s}\text{Tr}_{\phi}\Bigg[\,\sum_{\ts{I},\ts{J} \in \{\ts{A,B} \}} \int \dd V \dd V' \dd V'' \dd V''' \Big(\hat{h}_{\ts{i}}(\mf x)  \hat{h}_{\ts{c}}(\mf x') + \hat{h}_{\ts{c}}(\mf x)\hat{h}_{\ts{i}}(\mf x')\Big) \rhoh^{(0)}_{\ts{abc}} \nonumber \\[-0.4cm]
   &\phantom{==============}\times\Big(\hat{h}_{\ts{j}}(\mf x)  \hat{h}_{\ts{c}}(\mf x') + \hat{h}_{\ts{c}}(\mf x)\hat{h}_{\ts{j}}(\mf x')\Big)\theta(t-t') \theta(t''-t''') \Bigg]\! \ket{s} + \epsilon\, \mathcal{O}(\lambda^4)\;.
\end{align}
This can be rewritten as
\begin{align}\label{appAbis}
   \bra{s}\hat{U}^{(2)}\! \,\rhoh_{\ts{abc}}^{(0)}\, \hat{U}^{(2)\dagger}\ket{s}&=\lambda^4 \!\sum_{\ts{I,J} \in \{\ts{A,B} \}}\!\int \dd V \dd V' \dd V'' \dd V''' \Lambda_{\ts i}(\mf x)\Lambda_{\ts c}(\mf x') \Lambda_{\ts j}(\mf x'') \Lambda_{\ts c}(\mf x''') e^{\ii \Omega_{\ts c} (\tau_{\ts c}'-\tau_{\ts c}''')} e^{\ii (\Omega_{\ts i}\tau_\ts{i}- \Omega_\ts{j} \tau_\ts{j}'')} \nonumber \\
    &\phantom{======}\times \Big[ W(\mf x''', \mf x'', \mf x, \mf x')\theta(t-t') \theta(t'' - t''')\! +\! W(\mf x''', \mf x'', \mf x', \mf x)\theta(t'-t) \theta(t'' - t''') \nonumber \\[2mm]
    &\phantom{========} +\!W(\mf x'', \mf x''', \mf x, \mf x')\theta(t-t') \theta(t''' - t'')\! +\! W(\mf x''', \mf x'', \mf x, \mf x')\theta(t'-t) \theta(t''' - t'')\Big] \qquad \nonumber\\[2mm]
    &\phantom{======}\times
    (\ketbra{e_\ts{a}}{g_\ts{a}})^{\delta_{\ts{ai}}}(\ketbra{e_\ts{b}}{g_\ts{b}})^{\delta_{\ts{bi}}}\ketbra{g_{\ts a}}{g_{\ts a}} \otimes \ketbra{g_{\ts b}}{g_{\ts b}}(\ketbra{g_\ts{a}}{e_\ts{a}})^{\delta_{\ts{aj}}}(\ketbra{g_\ts{b}}{e_\ts{b}})^{\delta_{\ts{bj}}} + \epsilon\,\mathcal{O}(\lambda^4)\;,
\end{align}
where $\delta_{\ts k \ts l}$ is the Kronecker delta. To go from Eq.~\eqref{appA} to~\eqref{appAbis} we have performed changes of variables $\mf x \leftrightarrow \mf x'$, $\mf x'' \leftrightarrow \mf x'''$ in order to have $\hat{h}_{\ts i}$ evaluated at $\mf x$ and $\hat{h}_\ts{j}$ at $\mf x''$. Since the field is initially in its vacuum state, which is a quasifree state, the 4-point function can be expressed in terms of the two point function as
\begin{align}
W(\mathsf{x}_1,\mathsf{x}_2,\mathsf{x}_3,\mathsf{x}_4)&=W(\mathsf{x}_1,\mathsf{x}_2)W(\mathsf{x}_3,\mathsf{x}_4)+W(\mathsf{x}_1,\mathsf{x}_3)W(\mathsf{x}_2,\mathsf{x}_4)+W(\mathsf{x}_1,\mathsf{x}_4)W(\mathsf{x}_2,\mathsf{x}_3)\,. \nonumber
\end{align}
 Combining this with the relation $\theta(t-t') + \theta(t'-t) = 1$, the expression above can be rearranged as
\begin{align}
    \bra{s}\hat{U}^{(2)}\! \,\rhoh_{\ts{abc}}^{(0)}\, \hat{U}^{(2)\dagger} \ket{s} &= \mathcal{L}_{\ts{cc}}\Tilde{\mathcal{L}}_{\ts{aa}}\ketbra{e_{\ts a}}{e_{\ts a}} \otimes \ketbra{g_{\ts b}}{g_{\ts b}} + \mathcal{L}_{\ts{cc}}\Tilde{\mathcal{L}}_{\ts{bb}}\ketbra{g_{\ts a}}{g_{\ts a}} \otimes \ketbra{e_{\ts b}}{e_{\ts b}} \nonumber \\[1mm]
    &\phantom{=}+\mathcal{L}_{\ts{cc}}\Tilde{\mathcal{L}}_{\ts{ab}}\ketbra{e_{\ts a}}{g_{\ts a}} \otimes \ketbra{g_{\ts b}}{e_{\ts b}}+
    \mathcal{L}_{\ts{cc}}\Tilde{\mathcal{L}}_{\ts{ab}}^{*}\ketbra{g_{\ts a}}{e_{\ts a}} \otimes \ketbra{e_{\ts b}}{g_{\ts b}} +  \epsilon\mathcal{O}(\lambda^4)\,, \label{eq:A15}
\end{align}
where $\mathcal{L}_{\ts{cc}}$ is the excitation probability of detector $\ts{C}$, and $\Tilde{\mathcal{L}}_\ts{ij}$ is given by Eq.~\eqref{eq:TLij}. We now compute the fourth order terms left, namely \mbox{$\bra{s}\hat{U}^{(3)}\! \,\rhoh_{\ts{abc}}^{(0)}\, \hat{U}^{(1)\dagger}\ket{s}$} and \mbox{$\bra{s}\hat{U}^{(1)}\! \,\rhoh_{\ts{abc}}^{(0)}\, \hat{U}^{(3)\dagger}\ket{s}$}:
\begin{align}
    &\bra{s}\hat{U}^{(3)}\! \,\rhoh_{\ts{abc}}^{(0)}\, \hat{U}^{(1)\dagger} \ket{s} =\bra{s}\text{Tr}_{\phi}\Bigg[\sum_{\ts{I,J} \in \{\ts{A,B} \}} \int \dd V \dd V' \dd V'' \dd V''' \Big(\hat{h}_{\ts{c}}(\mf x)  \hat{h}_{\ts i}(\mf x')\hat{h}_{\ts j}(\mf x'')+ \hat{h}_{\ts i}(\mf x)  \hat{h}_{\ts{c}}(\mf x')\hat{h}_{\ts j}(\mf x'')\nonumber \\
    &\phantom{=====} + \hat{h}_{\ts i}(\mf x)  \hat{h}_{\ts j}(\mf x')\hat{h}_{\ts{c}}(\mf x'')\Big) \,\rhoh^{(0)}_{\ts{abc}}\, \hat{h}_{\ts c}(\mf x''')\theta(t-t') \theta(t'-t'') \Bigg]\! \ket{s} +\kappa_1\ketbra{g_{\ts a}}{g_{\ts a}} \otimes \ketbra{e_{\ts b}}{e_{\ts b}} + \epsilon\, \mathcal{O}(\lambda^4)\,,
\end{align}
where $\kappa_1$ is a local term of $\ts{C}$ given by 
\begin{align} 
    \kappa_1 = \bra{s}\text{Tr}_{\phi}\Bigg[\sum_{\ts{I,J} \in \{\ts{A,B} \}} \int \dd V \dd V' \dd V'' \dd V''' \hat{h}_{\ts{c}}(\mf x)  \hat{h}_{\ts i}(\mf x')\hat{h}_{\ts j}(\mf x'')\rhoh^{(0)}_{\ts{abc}}\, \hat{h}_{\ts c}(\mf x''')\theta(t-t') \theta(t'-t'') \Bigg]\! \ket{s}. 
\end{align}
 Finally,
\begin{align}
    &\bra{s}\hat{U}^{(3)}\! \,\rhoh_{\ts{abc}}^{(0)}\, \hat{U}^{(1)\dagger} \ket{s} + \bra{s}\hat{U}^{(1)}\! \,\rhoh_{\ts{abc}}^{(0)}\, \hat{U}^{(3)\dagger} \ket{s} = \mathcal{L}_{\ts{cc}}\Tilde{\mathcal{M}}_{\ts{ab}} \ketbra{e_{\ts a}}{g_{\ts a}} \otimes \ketbra{e_{\ts b}}{g_{\ts b}}+ \mathcal{L}_{\ts{cc}}\Tilde{\mathcal{M}}_{\ts{ab}}^* \ketbra{g_{\ts a}}{e_{\ts a}} \otimes \ketbra{g_{\ts b}}{e_{\ts b}}\nonumber \\ &\phantom{========} - \mathcal{L}_{\ts{cc}}\Tilde{\mathcal{L}}_{\ts{aa}} \ketbra{g_{\ts a}}{g_{\ts a}} \otimes \ketbra{g_{\ts b}}{g_{\ts b}} - \mathcal{L}_{\ts{cc}}\Tilde{\mathcal{L}}_{\ts{bb}} \ketbra{g_{\ts a}}{g_{\ts a}} \otimes \ketbra{g_{\ts b}}{g_{\ts b}} + \kappa_2 \ketbra{g_{\ts b}}{g_{\ts b}}\otimes \ketbra{g_{\ts a}}{g_{\ts a}} + \epsilon \, \mathcal{O}(\lambda^4)\,, \label{eq:A20}
\end{align}
with $\kappa_2$ given by
\begin{equation}
    \kappa_2 = \text{Tr}\Big[\bra{s}\hat{U}^{(3)}\! \,\rhoh_{\ts{abc}}^{(0)}\, \hat{U}^{(1)\dagger}\ket{s} + \bra{s}\hat{U}^{(1)}\! \,\rhoh_{\ts{abc}}^{(0)}\, \hat{U}^{(3)\dagger}\ket{s} \Big].
\end{equation}
The fourth order terms are obtained adding the contributions of Eqs.~\eqref{eq:A14},~\eqref{eq:A15}, and~\eqref{eq:A20}, yielding
\begin{equation} \label{eq:fourth}
     \bra{s}\rhoh_{\ts{abc}}^{(4)} \ket{s} = \kappa  \rhoh^{(0)}_{\ts{ab}} + \hat{\nu} + \epsilon\,\mathcal{O}(\lambda^4)\,,
\end{equation}
with $\hat{\nu}$ given by Eq.~\eqref{eq:nu} and
\begin{equation}\label{eq:kappa}
    \kappa = \kappa_1 + \kappa_2\,. 
\end{equation}
Finally, we add the terms proportional to $\lambda^0$, $\lambda^2$ and $\lambda^4$, which are given by Eqs.~\eqref{eq:zeroth},~\eqref{eq:second}, and~\eqref{eq:fourth}, respectively. This completes the proof for the expansion stated in Eq.~\eqref{eq:numerator}.

\section{Negativity} \label{appendix negativity}

In this Appendix we show that the negativity of a density operator of the form of Eq.~\eqref{eq:matrix1} is given by Eq.~\eqref{eq:negativity1}. The negativity of a two-qubit system is a function of the negative eigenvalues of the partially transposed density operator (partially transposing with respect to either of the subsystems $\ts{A}$ or $\ts{B}$, which yields the same eigenvalues). Concretely, we can write
\begin{equation} \label{eq:defNegativity}
    \mathcal{N}(\rhoh_{\ts{ab}}) = \sum_{j} \frac{|x_j|-x_j}{2},
\end{equation}
where $x_j$ are the eigenvalues of the partially transposed density matrix. Note that the expression above reduces to the absolute value of the sum of negative eigenvalues of the partially transposed density matrix. Given the density matrix in Eq.~\eqref{eq:matrix1}, its partial transpose with respect to subsystem $\ts{B}$ is given by
\begin{equation} \label{eq:ptBmatrix1}
     \left(
\begin{array}{cccc}
1-\lambda^2r_{22}-\lambda^2r_{33} & \Theta(\lambda^b) & \Theta(\lambda^b) & \lambda^2r_{32}^\ast \\
\Theta(\lambda^b) & \lambda^2 r_{22} & \lambda^2 r_{41}^* & 0 \\
\Theta(\lambda^b) & \lambda^2 r_{41} & \lambda^2 r_{33} & 0 \\
\lambda^{2} r_{32} & 0 & 0 & 0 \\
\end{array}\right) +\mathcal{O}(\lambda^3) \,.
\end{equation}
The characteristic polynomial of the matrix above reads
\begin{equation} \label{eq:charPol}
    P(x) = x^4 - x^3 + \left(r_{22} + r_{33}  + \mathcal{O}(\lambda^3)\right)x^2 + \left(|r_{41}|^2 - r_{22}r_{33}+ \mathcal{O}(\lambda^5)\right) + \mathcal{O}(\lambda^8)\,.
\end{equation}
In order to find the roots of the above expression, we need to follow a method that is consistent with the perturbative expansion, as in, e.g.,~\cite{PetarOld}. Assume that $x$ is a root of Eq.~\eqref{eq:charPol}. We expand it in $\lambda$ as
\begin{equation}
    x = x^{(0)} + x^{(1)} + x^{(2)} +  O(\lambda^3)\,,
\end{equation}
 where $x^{(i)}$ is proportional to $\lambda^i$. We impose $P(x) = 0$ and set the terms of order $i$ in $\lambda$ equal to zero, starting by $i = 0$ and increasing the order until we determine $x^{(0)}$, $x^{(1)}$ and $x^{(2)}$ up to the order we need. In the first step, for $i=0$, we obtain
\begin{equation}
    \big[x^{(0)}\big]^3 \,\big[x^{(0)}-1\big] = 0\,.
\end{equation}
Hence, either $x^{(0)} = 0$ or $x^{(0)} = 1$. For $i=1$, we get
\begin{equation}
    4\big[x^{(0)}\big]^3\, x^{(1)} - 3\big[x^{(0)}\big]^2 \,x^{(1)} = 0 \,.
\end{equation}
If $x^{(0)} = 0$, the equation above vanishes. If $x^{(0)} = 1$, the equation allows to find $x^{(1)} = 0$. For $i=2$, we obtain the equation
\begin{equation}
    4\big[ x^{(0)}\big]^3 \,x^{(2)} + 6 \big[x^{(0)}x^{(1)}\big]^2 - 3 \big[ x^{(0)}\big]^2\, x^{(2)}  - 3 \big[ x^{(1)}\big]^2\, x^{(0)} + x^{(0)}\,(r_{22} + r_{33}) = 0\,.
\end{equation}
If the zeroth order term is zero, i.e., $x^{(0)} = 0$, all the terms of the equation above also vanish. If $x^{(0)} = 1$ and $x^{(1)} = 0$, the equation above simplifies to $x^{(2)} +  (r_{22} + r_{33})= 0$, which allows us to find the second order term of the eigenvalue. Therefore, we have completely characterized one of the eigenvalues up to second order in perturbation theory:
\begin{equation}\label{req:E1}
    x_{1} = 1 - r_{22} - r_{33} + \mathcal{O}(\lambda^3) \,.
\end{equation}
Henceforth, we can restrict to the case in which the zeroth order term vanishes, $x^{(0)} = 0$. For $i=3$, we obtain $\big[x^{(1)}\big]^3 = 0$,
which implies $x^{(1)} = 0$. Consequently, the remaining eigenvalues are quadratic in the coupling strength, $x = x^{(2)} + O(\lambda^3)$. There are no terms in the equation $P(x) = 0$ proportional to $\lambda^4$ or $\lambda^5$. For sixth order, we get
\begin{equation}
    -\big[x^{(2)}\big]^3 + (r_{22} + r_{33})\big[x^{(2)} \big]^2  + (|r_{41}|^2 - r_{22}r_{33}) x^{(2)} = 0 \,.
\end{equation}
Hence, $x^{(2)}$ is a root of the previous expression. The solutions are 
\begin{align}
    x^{(2)}_2 &= 0 \,, \\
    x^{(2)}_3 &= \frac{(r_{22} + r_{33}) + \sqrt{(r_{22} + r_{33})^2 + 4(|r_{41}|^2 - r_{22}r_{33})}}{2} \,, \\
    x^{(2)}_4 &= \frac{(r_{22} + r_{33}) - \sqrt{(r_{22} + r_{33})^2 + 4(|r_{41}|^2 - r_{22}r_{33})}}{2} \,.
\end{align}
Therefore, the eigenvalues left read 
\begin{align}
    &x_2 =  0 + \mathcal{O}(\lambda^3)\,, \label{req:E2} \\
    &x_{3} =  \frac{(r_{22} + r_{33}) + \sqrt{(r_{22} + r_{33})^2 + 4(|r_{41}|^2 - r_{22}r_{33})}}{2}  +\mathcal{O}(\lambda^3)\,, \label{req:E3} \\
    &x_{4} =  \frac{(r_{22} + r_{33}) - \sqrt{(r_{22} + r_{33})^2 + 4(|r_{41}|^2 - r_{22}r_{33})}}{2} +\mathcal{O}(\lambda^3)\,. \label{req:E4}
\end{align}
The negativity is derived from the negative eigenvalues of the partially transposed density operator. The eigenvalue $x_1$ is always positive, since we are assuming that $\lambda$ is a small parameter. The eigenvalue $x_2$ is zero (and therefore non-negative) to second order in perturbation theory. The eigenvalue $x_3$ is positive since it is a sum of positive terms. Therefore, the only eigenvalue that can be negative (to second order) is $x_4$. Thus the negativity is
\begin{align}
    \mathcal{N}_s &= 
    \text{max}\left(0\,\sqrt{\frac{(r_{22} + r_{33})^2}{4}+ |r_{41}|^2 - r_{22}r_{33}}\, -\,\frac{r_{22} + r_{33}}{2} +\mathcal{O}(\lambda^3)\right) \nonumber \\
    &= \text{max}\left(0\,,\sqrt{\frac{(r_{22} - r_{33})^2}{4}+ |r_{41}|^2}\, -\,\frac{r_{22} + r_{33}}{2} +\mathcal{O}(\lambda^3)\right). \label{req:negativityAlgebraic}
\end{align}

\twocolumngrid

\bibliography{references}

\begin{thebibliography}{61}%
\makeatletter
\providecommand \@ifxundefined [1]{%
 \@ifx{#1\undefined}
}%
\providecommand \@ifnum [1]{%
 \ifnum #1\expandafter \@firstoftwo
 \else \expandafter \@secondoftwo
 \fi
}%
\providecommand \@ifx [1]{%
 \ifx #1\expandafter \@firstoftwo
 \else \expandafter \@secondoftwo
 \fi
}%
\providecommand \natexlab [1]{#1}%
\providecommand \enquote  [1]{``#1''}%
\providecommand \bibnamefont  [1]{#1}%
\providecommand \bibfnamefont [1]{#1}%
\providecommand \citenamefont [1]{#1}%
\providecommand \href@noop [0]{\@secondoftwo}%
\providecommand \href [0]{\begingroup \@sanitize@url \@href}%
\providecommand \@href[1]{\@@startlink{#1}\@@href}%
\providecommand \@@href[1]{\endgroup#1\@@endlink}%
\providecommand \@sanitize@url [0]{\catcode `\\12\catcode `\$12\catcode
  `\&12\catcode `\#12\catcode `\^12\catcode `\_12\catcode `\%12\relax}%
\providecommand \@@startlink[1]{}%
\providecommand \@@endlink[0]{}%
\providecommand \url  [0]{\begingroup\@sanitize@url \@url }%
\providecommand \@url [1]{\endgroup\@href {#1}{\urlprefix }}%
\providecommand \urlprefix  [0]{URL }%
\providecommand \Eprint [0]{\href }%
\providecommand \doibase [0]{https://doi.org/}%
\providecommand \selectlanguage [0]{\@gobble}%
\providecommand \bibinfo  [0]{\@secondoftwo}%
\providecommand \bibfield  [0]{\@secondoftwo}%
\providecommand \translation [1]{[#1]}%
\providecommand \BibitemOpen [0]{}%
\providecommand \bibitemStop [0]{}%
\providecommand \bibitemNoStop [0]{.\EOS\space}%
\providecommand \EOS [0]{\spacefactor3000\relax}%
\providecommand \BibitemShut  [1]{\csname bibitem#1\endcsname}%
\let\auto@bib@innerbib\@empty
\bibitem [{\citenamefont {Hotta}(2009)}]{Hotta2009}%
  \BibitemOpen
  \bibfield  {author} {\bibinfo {author} {\bibfnamefont {M.}~\bibnamefont
  {Hotta}},\ }\bibfield  {title} {\bibinfo {title} {Quantum energy
  teleportation in spin chain systems},\ }\href
  {https://doi.org/10.1143/JPSJ.78.034001} {\bibfield  {journal} {\bibinfo
  {journal} {J. Phys. Soc. Japan}\ }\textbf {\bibinfo {volume} {78}},\ \bibinfo
  {pages} {034001} (\bibinfo {year} {2009})}\BibitemShut {NoStop}%
\bibitem [{\citenamefont {Hotta}(2011)}]{Hotta2011}%
  \BibitemOpen
  \bibfield  {author} {\bibinfo {author} {\bibfnamefont {M.}~\bibnamefont
  {Hotta}},\ }\href@noop {} {\bibinfo {title} {{Q}uantum {E}nergy
  {T}eleportation: {A}n {I}ntroductory {R}eview}} (\bibinfo {year} {2011}),\
  \Eprint {https://arxiv.org/abs/1101.3954} {arXiv:1101.3954 [quant-ph]}
  \BibitemShut {NoStop}%
\bibitem [{\citenamefont {Preskill}(1992)}]{Preskill1992}%
  \BibitemOpen
  \bibfield  {author} {\bibinfo {author} {\bibfnamefont {J.}~\bibnamefont
  {Preskill}},\ }\href@noop {} {\bibinfo {title} {Do {B}lack {H}oles {D}estroy
  {I}nformation?}} (\bibinfo {year} {1992}),\ \Eprint
  {https://arxiv.org/abs/hep-th/9209058} {arXiv:hep-th/9209058} \BibitemShut
  {NoStop}%
\bibitem [{\citenamefont {Klebanov}\ \emph {et~al.}(2008)\citenamefont
  {Klebanov}, \citenamefont {Kutasov},\ and\ \citenamefont
  {Murugan}}]{Klebanov2008}%
  \BibitemOpen
  \bibfield  {author} {\bibinfo {author} {\bibfnamefont {I.~R.}\ \bibnamefont
  {Klebanov}}, \bibinfo {author} {\bibfnamefont {D.}~\bibnamefont {Kutasov}},\
  and\ \bibinfo {author} {\bibfnamefont {A.}~\bibnamefont {Murugan}},\
  }\bibfield  {title} {\bibinfo {title} {Entanglement as a probe of
  confinement},\ }\href
  {https://doi.org/https://doi.org/10.1016/j.nuclphysb.2007.12.017} {\bibfield
  {journal} {\bibinfo  {journal} {Nucl. Phys.}\ }\textbf {\bibinfo {volume}
  {B796}},\ \bibinfo {pages} {274} (\bibinfo {year} {2008})}\BibitemShut
  {NoStop}%
\bibitem [{\citenamefont {Jokela}\ and\ \citenamefont
  {Subils}(2021)}]{Jokela2021}%
  \BibitemOpen
  \bibfield  {author} {\bibinfo {author} {\bibfnamefont {N.}~\bibnamefont
  {Jokela}}\ and\ \bibinfo {author} {\bibfnamefont {J.~G.}\ \bibnamefont
  {Subils}},\ }\bibfield  {title} {\bibinfo {title} {Is entanglement a probe of
  confinement?},\ }\href {https://doi.org/10.1007/JHEP02(2021)147} {\bibfield
  {journal} {\bibinfo  {journal} {J. High Energy Phys.}\ }\textbf {\bibinfo
  {volume} {2021}}\bibinfo  {number} { (2)},\ \bibinfo {pages}
  {147}}\BibitemShut {NoStop}%
\bibitem [{\citenamefont {Reeh}\ and\ \citenamefont
  {Schlieder}(1961)}]{ReehSchlieder}%
  \BibitemOpen
\bibfield  {number} {  }\bibfield  {author} {\bibinfo {author} {\bibfnamefont
  {H.}~\bibnamefont {Reeh}}\ and\ \bibinfo {author} {\bibfnamefont
  {S.}~\bibnamefont {Schlieder}},\ }\bibfield  {title} {\bibinfo {title}
  {Bemerkungen zur unit\"{a}r\"{a}quivalenz von lorentzinvarianten feldern},\
  }\href {https://doi.org/10.1007/BF02787889} {\bibfield  {journal} {\bibinfo
  {journal} {Nuovo Cimento}\ }\textbf {\bibinfo {volume} {22}},\ \bibinfo
  {pages} {1051} (\bibinfo {year} {1961})}\BibitemShut {NoStop}%
\bibitem [{\citenamefont {Schlieder}(1968)}]{Schlieder1968}%
  \BibitemOpen
  \bibfield  {author} {\bibinfo {author} {\bibfnamefont {S.}~\bibnamefont
  {Schlieder}},\ }\bibfield  {title} {\bibinfo {title} {Einige {B}emerkungen
  zur {Z}ustands\"{a}nderung von relativistischen quantenmechanischen
  {S}ystemen durch {M}essungen und zur {L}okalit\"{a}tsforderung},\ }\href
  {https://doi.org/10.1007/BF01646663} {\bibfield  {journal} {\bibinfo
  {journal} {Commun. Math. Phys.}\ }\textbf {\bibinfo {volume} {7}},\ \bibinfo
  {pages} {305} (\bibinfo {year} {1968})}\BibitemShut {NoStop}%
\bibitem [{\citenamefont {Sorkin}(1993)}]{Sorkin}%
  \BibitemOpen
  \bibfield  {author} {\bibinfo {author} {\bibfnamefont {R.~D.}\ \bibnamefont
  {Sorkin}},\ }\href@noop {} {\bibinfo {title} {Impossible {M}easurements on
  {Q}uantum {F}ields}} (\bibinfo {year} {1993}),\ \Eprint
  {https://arxiv.org/abs/gr-qc/9302018} {arXiv:gr-qc/9302018 [gr-qc]}
  \BibitemShut {NoStop}%
\bibitem [{\citenamefont {Borsten}\ \emph {et~al.}(2021)\citenamefont
  {Borsten}, \citenamefont {Jubb},\ and\ \citenamefont {Kells}}]{Borsten2021}%
  \BibitemOpen
  \bibfield  {author} {\bibinfo {author} {\bibfnamefont {L.}~\bibnamefont
  {Borsten}}, \bibinfo {author} {\bibfnamefont {I.}~\bibnamefont {Jubb}},\ and\
  \bibinfo {author} {\bibfnamefont {G.}~\bibnamefont {Kells}},\ }\bibfield
  {title} {\bibinfo {title} {Impossible measurements revisited},\ }\href
  {https://doi.org/10.1103/PhysRevD.104.025012} {\bibfield  {journal} {\bibinfo
   {journal} {Phys. Rev. D}\ }\textbf {\bibinfo {volume} {104}},\ \bibinfo
  {pages} {025012} (\bibinfo {year} {2021})}\BibitemShut {NoStop}%
\bibitem [{\citenamefont {Summers}\ and\ \citenamefont
  {Werner}(1985)}]{SUMMERS}%
  \BibitemOpen
  \bibfield  {author} {\bibinfo {author} {\bibfnamefont {S.~J.}\ \bibnamefont
  {Summers}}\ and\ \bibinfo {author} {\bibfnamefont {R.}~\bibnamefont
  {Werner}},\ }\bibfield  {title} {\bibinfo {title} {The vacuum violates
  {B}ell's inequalities},\ }\href
  {https://doi.org/https://doi.org/10.1016/0375-9601(85)90093-3} {\bibfield
  {journal} {\bibinfo  {journal} {Phys. Lett. A}\ }\textbf {\bibinfo {volume}
  {110}},\ \bibinfo {pages} {257} (\bibinfo {year} {1985})}\BibitemShut
  {NoStop}%
\bibitem [{\citenamefont {Higuchi}\ \emph {et~al.}(2017)\citenamefont
  {Higuchi}, \citenamefont {Iso}, \citenamefont {Ueda},\ and\ \citenamefont
  {Yamamoto}}]{vacuumEntanglement}%
  \BibitemOpen
  \bibfield  {author} {\bibinfo {author} {\bibfnamefont {A.}~\bibnamefont
  {Higuchi}}, \bibinfo {author} {\bibfnamefont {S.}~\bibnamefont {Iso}},
  \bibinfo {author} {\bibfnamefont {K.}~\bibnamefont {Ueda}},\ and\ \bibinfo
  {author} {\bibfnamefont {K.}~\bibnamefont {Yamamoto}},\ }\bibfield  {title}
  {\bibinfo {title} {Entanglement of the vacuum between left, right, future,
  and past: The origin of entanglement-induced quantum radiation},\ }\href
  {https://doi.org/10.1103/PhysRevD.96.083531} {\bibfield  {journal} {\bibinfo
  {journal} {Phys. Rev. D}\ }\textbf {\bibinfo {volume} {96}},\ \bibinfo
  {pages} {083531} (\bibinfo {year} {2017})}\BibitemShut {NoStop}%
\bibitem [{\citenamefont {Bombelli}\ \emph {et~al.}(1986)\citenamefont
  {Bombelli}, \citenamefont {Koul}, \citenamefont {Lee},\ and\ \citenamefont
  {Sorkin}}]{Bombelli1986}%
  \BibitemOpen
  \bibfield  {author} {\bibinfo {author} {\bibfnamefont {L.}~\bibnamefont
  {Bombelli}}, \bibinfo {author} {\bibfnamefont {R.~K.}\ \bibnamefont {Koul}},
  \bibinfo {author} {\bibfnamefont {J.}~\bibnamefont {Lee}},\ and\ \bibinfo
  {author} {\bibfnamefont {R.~D.}\ \bibnamefont {Sorkin}},\ }\bibfield  {title}
  {\bibinfo {title} {Quantum source of entropy for black holes},\ }\href
  {https://doi.org/10.1103/PhysRevD.34.373} {\bibfield  {journal} {\bibinfo
  {journal} {Phys. Rev. D}\ }\textbf {\bibinfo {volume} {34}},\ \bibinfo
  {pages} {373} (\bibinfo {year} {1986})}\BibitemShut {NoStop}%
\bibitem [{\citenamefont {Witten}(2018)}]{witten}%
  \BibitemOpen
  \bibfield  {author} {\bibinfo {author} {\bibfnamefont {E.}~\bibnamefont
  {Witten}},\ }\bibfield  {title} {\bibinfo {title} {{APS} {M}edal for
  {E}xceptional {A}chievement in {R}esearch: {I}nvited article on entanglement
  properties of quantum field theory},\ }\href
  {https://doi.org/10.1103/RevModPhys.90.045003} {\bibfield  {journal}
  {\bibinfo  {journal} {Rev. Mod. Phys.}\ }\textbf {\bibinfo {volume} {90}},\
  \bibinfo {pages} {045003} (\bibinfo {year} {2018})}\BibitemShut {NoStop}%
\bibitem [{\citenamefont {Valentini}(1991)}]{Valentini1991}%
  \BibitemOpen
  \bibfield  {author} {\bibinfo {author} {\bibfnamefont {A.}~\bibnamefont
  {Valentini}},\ }\bibfield  {title} {\bibinfo {title} {Non-local correlations
  in quantum electrodynamics},\ }\href
  {https://doi.org/http://dx.doi.org/10.1016/0375-9601(91)90952-5} {\bibfield
  {journal} {\bibinfo  {journal} {Phys. Lett.}\ }\textbf {\bibinfo {volume}
  {153A}},\ \bibinfo {pages} {321 } (\bibinfo {year} {1991})}\BibitemShut
  {NoStop}%
\bibitem [{\citenamefont {Reznik}(2003)}]{Reznik2003}%
  \BibitemOpen
  \bibfield  {author} {\bibinfo {author} {\bibfnamefont {B.}~\bibnamefont
  {Reznik}},\ }\bibfield  {title} {\bibinfo {title} {Entanglement from the
  {V}acuum},\ }\href {https://doi.org/10.1023/A:1022875910744} {\bibfield
  {journal} {\bibinfo  {journal} {Found. Phys.}\ }\textbf {\bibinfo {volume}
  {33}},\ \bibinfo {pages} {167} (\bibinfo {year} {2003})}\BibitemShut
  {NoStop}%
\bibitem [{\citenamefont {Reznik}\ \emph {et~al.}(2005)\citenamefont {Reznik},
  \citenamefont {Retzker},\ and\ \citenamefont {Silman}}]{Reznik1}%
  \BibitemOpen
  \bibfield  {author} {\bibinfo {author} {\bibfnamefont {B.}~\bibnamefont
  {Reznik}}, \bibinfo {author} {\bibfnamefont {A.}~\bibnamefont {Retzker}},\
  and\ \bibinfo {author} {\bibfnamefont {J.}~\bibnamefont {Silman}},\
  }\bibfield  {title} {\bibinfo {title} {Violating {B}ell's inequalities in
  vacuum},\ }\href {http://link.aps.org/abstract/PRA/v71/e042104} {\bibfield
  {journal} {\bibinfo  {journal} {Phys. Rev. A}\ }\textbf {\bibinfo {volume}
  {71}},\ \bibinfo {eid} {042104} (\bibinfo {year} {2005})}\BibitemShut
  {NoStop}%
\bibitem [{\citenamefont {Silman}\ and\ \citenamefont
  {Reznik}(2007)}]{Reznik2}%
  \BibitemOpen
  \bibfield  {author} {\bibinfo {author} {\bibfnamefont {J.}~\bibnamefont
  {Silman}}\ and\ \bibinfo {author} {\bibfnamefont {B.}~\bibnamefont
  {Reznik}},\ }\bibfield  {title} {\bibinfo {title} {Long-range entanglement in
  the {D}irac vacuum},\ }\href {https://doi.org/10.1103/PhysRevA.75.052307}
  {\bibfield  {journal} {\bibinfo  {journal} {Phys. Rev. A}\ }\textbf {\bibinfo
  {volume} {75}},\ \bibinfo {pages} {052307} (\bibinfo {year}
  {2007})}\BibitemShut {NoStop}%
\bibitem [{\citenamefont {Salton}\ \emph
  {et~al.}(2015{\natexlab{a}})\citenamefont {Salton}, \citenamefont {Mann},\
  and\ \citenamefont {Menicucci}}]{Salton:2014jaa}%
  \BibitemOpen
  \bibfield  {author} {\bibinfo {author} {\bibfnamefont {G.}~\bibnamefont
  {Salton}}, \bibinfo {author} {\bibfnamefont {R.~B.}\ \bibnamefont {Mann}},\
  and\ \bibinfo {author} {\bibfnamefont {N.~C.}\ \bibnamefont {Menicucci}},\
  }\bibfield  {title} {\bibinfo {title} {{Acceleration-assisted entanglement
  harvesting and rangefinding}},\ }\href
  {https://doi.org/10.1088/1367-2630/17/3/035001} {\bibfield  {journal}
  {\bibinfo  {journal} {New J. Phys.}\ }\textbf {\bibinfo {volume} {17}},\
  \bibinfo {pages} {035001} (\bibinfo {year} {2015}{\natexlab{a}})}\BibitemShut
  {NoStop}%
\bibitem [{\citenamefont {Steeg}\ and\ \citenamefont {Menicucci}(2009)}]{Nick}%
  \BibitemOpen
  \bibfield  {author} {\bibinfo {author} {\bibfnamefont {G.~V.}\ \bibnamefont
  {Steeg}}\ and\ \bibinfo {author} {\bibfnamefont {N.~C.}\ \bibnamefont
  {Menicucci}},\ }\bibfield  {title} {\bibinfo {title} {Entangling power of an
  expanding universe},\ }\href {https://doi.org/10.1103/PhysRevD.79.044027}
  {\bibfield  {journal} {\bibinfo  {journal} {Phys. Rev. D}\ }\textbf {\bibinfo
  {volume} {79}},\ \bibinfo {pages} {044027} (\bibinfo {year}
  {2009})}\BibitemShut {NoStop}%
\bibitem [{\citenamefont {Cliche}\ and\ \citenamefont
  {Kempf}(2011)}]{Cliche2011}%
  \BibitemOpen
  \bibfield  {author} {\bibinfo {author} {\bibfnamefont {M.}~\bibnamefont
  {Cliche}}\ and\ \bibinfo {author} {\bibfnamefont {A.}~\bibnamefont {Kempf}},\
  }\bibfield  {title} {\bibinfo {title} {Vacuum entanglement enhancement by a
  weak gravitational field},\ }\href
  {https://doi.org/10.1103/PhysRevD.83.045019} {\bibfield  {journal} {\bibinfo
  {journal} {Phys. Rev. D}\ }\textbf {\bibinfo {volume} {83}},\ \bibinfo
  {pages} {045019} (\bibinfo {year} {2011})}\BibitemShut {NoStop}%
\bibitem [{\citenamefont {Pozas-Kerstjens}\ and\ \citenamefont
  {Mart\'{i}n-Mart\'{i}nez}(2015)}]{Pozas-Kerstjens:2015}%
  \BibitemOpen
  \bibfield  {author} {\bibinfo {author} {\bibfnamefont {A.}~\bibnamefont
  {Pozas-Kerstjens}}\ and\ \bibinfo {author} {\bibfnamefont {E.}~\bibnamefont
  {Mart\'{i}n-Mart\'{i}nez}},\ }\bibfield  {title} {\bibinfo {title}
  {Harvesting correlations from the quantum vacuum},\ }\href
  {https://doi.org/10.1103/PhysRevD.92.064042} {\bibfield  {journal} {\bibinfo
  {journal} {Phys. Rev. D}\ }\textbf {\bibinfo {volume} {92}},\ \bibinfo
  {pages} {064042} (\bibinfo {year} {2015})}\BibitemShut {NoStop}%
\bibitem [{\citenamefont {Pozas-Kerstjens}\ and\ \citenamefont
  {Mart\'{i}n-Mart\'{i}nez}(2016)}]{Pozas2016}%
  \BibitemOpen
  \bibfield  {author} {\bibinfo {author} {\bibfnamefont {A.}~\bibnamefont
  {Pozas-Kerstjens}}\ and\ \bibinfo {author} {\bibfnamefont {E.}~\bibnamefont
  {Mart\'{i}n-Mart\'{i}nez}},\ }\bibfield  {title} {\bibinfo {title}
  {Entanglement harvesting from the electromagnetic vacuum with hydrogenlike
  atoms},\ }\href {https://doi.org/10.1103/PhysRevD.94.064074} {\bibfield
  {journal} {\bibinfo  {journal} {Phys. Rev. D}\ }\textbf {\bibinfo {volume}
  {94}},\ \bibinfo {pages} {064074} (\bibinfo {year} {2016})}\BibitemShut
  {NoStop}%
\bibitem [{\citenamefont {Mart\'{\i}n-Mart\'{\i}nez}\ \emph
  {et~al.}(2016)\citenamefont {Mart\'{\i}n-Mart\'{\i}nez}, \citenamefont
  {Smith},\ and\ \citenamefont {Terno}}]{topology}%
  \BibitemOpen
  \bibfield  {author} {\bibinfo {author} {\bibfnamefont {E.}~\bibnamefont
  {Mart\'{\i}n-Mart\'{\i}nez}}, \bibinfo {author} {\bibfnamefont {A.~R.~H.}\
  \bibnamefont {Smith}},\ and\ \bibinfo {author} {\bibfnamefont {D.~R.}\
  \bibnamefont {Terno}},\ }\bibfield  {title} {\bibinfo {title} {Spacetime
  structure and vacuum entanglement},\ }\href
  {https://doi.org/10.1103/PhysRevD.93.044001} {\bibfield  {journal} {\bibinfo
  {journal} {Phys. Rev. D}\ }\textbf {\bibinfo {volume} {93}},\ \bibinfo
  {pages} {044001} (\bibinfo {year} {2016})}\BibitemShut {NoStop}%
\bibitem [{\citenamefont {Simidzija}\ and\ \citenamefont
  {Mart\'{\i}n-Mart\'{\i}nez}(2017)}]{PetarOld}%
  \BibitemOpen
  \bibfield  {author} {\bibinfo {author} {\bibfnamefont {P.}~\bibnamefont
  {Simidzija}}\ and\ \bibinfo {author} {\bibfnamefont {E.}~\bibnamefont
  {Mart\'{\i}n-Mart\'{\i}nez}},\ }\bibfield  {title} {\bibinfo {title} {All
  coherent field states entangle equally},\ }\href
  {https://doi.org/10.1103/PhysRevD.96.025020} {\bibfield  {journal} {\bibinfo
  {journal} {Phys. Rev. D}\ }\textbf {\bibinfo {volume} {96}},\ \bibinfo
  {pages} {025020} (\bibinfo {year} {2017})}\BibitemShut {NoStop}%
\bibitem [{\citenamefont {Simidzija}\ and\ \citenamefont
  {Mart\'{i}n-Mart\'{i}nez}(2018)}]{Petar}%
  \BibitemOpen
  \bibfield  {author} {\bibinfo {author} {\bibfnamefont {P.}~\bibnamefont
  {Simidzija}}\ and\ \bibinfo {author} {\bibfnamefont {E.}~\bibnamefont
  {Mart\'{i}n-Mart\'{i}nez}},\ }\bibfield  {title} {\bibinfo {title}
  {Harvesting correlations from thermal and squeezed coherent states},\ }\href
  {https://doi.org/10.1103/PhysRevD.98.085007} {\bibfield  {journal} {\bibinfo
  {journal} {Phys. Rev. D}\ }\textbf {\bibinfo {volume} {98}},\ \bibinfo
  {pages} {085007} (\bibinfo {year} {2018})}\BibitemShut {NoStop}%
\bibitem [{\citenamefont {Ng}\ \emph {et~al.}(2018)\citenamefont {Ng},
  \citenamefont {Mann},\ and\ \citenamefont {Mart\'{i}n-Mart\'{i}nez}}]{Ng2}%
  \BibitemOpen
  \bibfield  {author} {\bibinfo {author} {\bibfnamefont {K.~K.}\ \bibnamefont
  {Ng}}, \bibinfo {author} {\bibfnamefont {R.~B.}\ \bibnamefont {Mann}},\ and\
  \bibinfo {author} {\bibfnamefont {E.}~\bibnamefont
  {Mart\'{i}n-Mart\'{i}nez}},\ }\bibfield  {title} {\bibinfo {title}
  {Unruh-{D}e{W}itt detectors and entanglement: The anti--de {S}itter space},\
  }\href {https://doi.org/10.1103/PhysRevD.98.125005} {\bibfield  {journal}
  {\bibinfo  {journal} {Phys. Rev. D}\ }\textbf {\bibinfo {volume} {98}},\
  \bibinfo {pages} {125005} (\bibinfo {year} {2018})}\BibitemShut {NoStop}%
\bibitem [{\citenamefont {Henderson}\ \emph {et~al.}(2018)\citenamefont
  {Henderson}, \citenamefont {Hennigar}, \citenamefont {Mann}, \citenamefont
  {Smith},\ and\ \citenamefont {Zhang}}]{Henderson2018}%
  \BibitemOpen
  \bibfield  {author} {\bibinfo {author} {\bibfnamefont {L.~J.}\ \bibnamefont
  {Henderson}}, \bibinfo {author} {\bibfnamefont {R.~A.}\ \bibnamefont
  {Hennigar}}, \bibinfo {author} {\bibfnamefont {R.~B.}\ \bibnamefont {Mann}},
  \bibinfo {author} {\bibfnamefont {A.~R.~H.}\ \bibnamefont {Smith}},\ and\
  \bibinfo {author} {\bibfnamefont {J.}~\bibnamefont {Zhang}},\ }\bibfield
  {title} {\bibinfo {title} {Harvesting entanglement from the black hole
  vacuum},\ }\href {https://doi.org/10.1088/1361-6382/aae27e} {\bibfield
  {journal} {\bibinfo  {journal} {Class. Quantum Gravity}\ }\textbf {\bibinfo
  {volume} {35}},\ \bibinfo {pages} {21LT02} (\bibinfo {year}
  {2018})}\BibitemShut {NoStop}%
\bibitem [{\citenamefont {Henderson}\ \emph {et~al.}(2019)\citenamefont
  {Henderson}, \citenamefont {Hennigar}, \citenamefont {Mann}, \citenamefont
  {Smith},\ and\ \citenamefont {Zhang}}]{Henderson2019}%
  \BibitemOpen
  \bibfield  {author} {\bibinfo {author} {\bibfnamefont {L.~J.}\ \bibnamefont
  {Henderson}}, \bibinfo {author} {\bibfnamefont {R.~A.}\ \bibnamefont
  {Hennigar}}, \bibinfo {author} {\bibfnamefont {R.~B.}\ \bibnamefont {Mann}},
  \bibinfo {author} {\bibfnamefont {A.~R.~H.}\ \bibnamefont {Smith}},\ and\
  \bibinfo {author} {\bibfnamefont {J.}~\bibnamefont {Zhang}},\ }\bibfield
  {title} {\bibinfo {title} {Entangling detectors in anti-de {S}itter space},\
  }\href {https://doi.org/10.1007/JHEP05(2019)178} {\bibfield  {journal}
  {\bibinfo  {journal} {J. High Energy Phys.}\ }\textbf {\bibinfo {volume}
  {2019}}\bibinfo  {number} { (5)},\ \bibinfo {pages} {178}}\BibitemShut
  {NoStop}%
\bibitem [{\citenamefont {Henderson}\ and\ \citenamefont
  {Menicucci}(2020)}]{Henderson2020}%
  \BibitemOpen
\bibfield  {number} {  }\bibfield  {author} {\bibinfo {author} {\bibfnamefont
  {L.~J.}\ \bibnamefont {Henderson}}\ and\ \bibinfo {author} {\bibfnamefont
  {N.~C.}\ \bibnamefont {Menicucci}},\ }\bibfield  {title} {\bibinfo {title}
  {Bandlimited entanglement harvesting},\ }\href
  {https://doi.org/10.1103/PhysRevD.102.125026} {\bibfield  {journal} {\bibinfo
   {journal} {Phys. Rev. D}\ }\textbf {\bibinfo {volume} {102}},\ \bibinfo
  {pages} {125026} (\bibinfo {year} {2020})}\BibitemShut {NoStop}%
\bibitem [{\citenamefont {Tjoa}\ and\ \citenamefont
  {Mann}(2020)}]{Erickson2020}%
  \BibitemOpen
  \bibfield  {author} {\bibinfo {author} {\bibfnamefont {E.}~\bibnamefont
  {Tjoa}}\ and\ \bibinfo {author} {\bibfnamefont {R.~B.}\ \bibnamefont
  {Mann}},\ }\bibfield  {title} {\bibinfo {title} {Harvesting correlations in
  {S}chwarzschild and collapsing shell spacetimes},\ }\href
  {https://doi.org/10.1007/JHEP08(2020)155} {\bibfield  {journal} {\bibinfo
  {journal} {J. High Energy Phys.}\ }\textbf {\bibinfo {volume} {2020}}\bibinfo
   {number} { (8)},\ \bibinfo {pages} {155}}\BibitemShut {NoStop}%
\bibitem [{\citenamefont {Tjoa}\ and\ \citenamefont
  {Mart\'{\i}n-Mart\'{\i}nez}(2021)}]{ericksonNew}%
  \BibitemOpen
\bibfield  {number} {  }\bibfield  {author} {\bibinfo {author} {\bibfnamefont
  {E.}~\bibnamefont {Tjoa}}\ and\ \bibinfo {author} {\bibfnamefont
  {E.}~\bibnamefont {Mart\'{\i}n-Mart\'{\i}nez}},\ }\bibfield  {title}
  {\bibinfo {title} {When entanglement harvesting is not really harvesting},\
  }\href {https://doi.org/10.1103/PhysRevD.104.125005} {\bibfield  {journal}
  {\bibinfo  {journal} {Phys. Rev. D}\ }\textbf {\bibinfo {volume} {104}},\
  \bibinfo {pages} {125005} (\bibinfo {year} {2021})}\BibitemShut {NoStop}%
\bibitem [{\citenamefont {Foo}\ \emph {et~al.}(2021)\citenamefont {Foo},
  \citenamefont {Mann},\ and\ \citenamefont {Zych}}]{Foo2021}%
  \BibitemOpen
  \bibfield  {author} {\bibinfo {author} {\bibfnamefont {J.}~\bibnamefont
  {Foo}}, \bibinfo {author} {\bibfnamefont {R.~B.}\ \bibnamefont {Mann}},\ and\
  \bibinfo {author} {\bibfnamefont {M.}~\bibnamefont {Zych}},\ }\bibfield
  {title} {\bibinfo {title} {Entanglement amplification between superposed
  detectors in flat and curved spacetimes},\ }\href
  {https://doi.org/10.1103/PhysRevD.103.065013} {\bibfield  {journal} {\bibinfo
   {journal} {Phys. Rev. D}\ }\textbf {\bibinfo {volume} {103}},\ \bibinfo
  {pages} {065013} (\bibinfo {year} {2021})}\BibitemShut {NoStop}%
\bibitem [{\citenamefont {Perche}\ \emph {et~al.}(2022)\citenamefont {Perche},
  \citenamefont {Lima},\ and\ \citenamefont
  {Mart\'{\i}n-Mart\'{\i}nez}}]{carol}%
  \BibitemOpen
  \bibfield  {author} {\bibinfo {author} {\bibfnamefont {T.~R.}\ \bibnamefont
  {Perche}}, \bibinfo {author} {\bibfnamefont {C.}~\bibnamefont {Lima}},\ and\
  \bibinfo {author} {\bibfnamefont {E.}~\bibnamefont
  {Mart\'{\i}n-Mart\'{\i}nez}},\ }\bibfield  {title} {\bibinfo {title}
  {Harvesting entanglement from complex scalar and fermionic fields with
  linearly coupled particle detectors},\ }\href
  {https://doi.org/10.1103/PhysRevD.105.065016} {\bibfield  {journal} {\bibinfo
   {journal} {Phys. Rev. D}\ }\textbf {\bibinfo {volume} {105}},\ \bibinfo
  {pages} {065016} (\bibinfo {year} {2022})}\BibitemShut {NoStop}%
\bibitem [{\citenamefont {Sahu}\ \emph {et~al.}(2022)\citenamefont {Sahu},
  \citenamefont {Melgarejo-Lermas},\ and\ \citenamefont
  {Mart\'{\i}n-Mart\'{\i}nez}}]{Sahu2022}%
  \BibitemOpen
  \bibfield  {author} {\bibinfo {author} {\bibfnamefont {A.}~\bibnamefont
  {Sahu}}, \bibinfo {author} {\bibfnamefont {I.}~\bibnamefont
  {Melgarejo-Lermas}},\ and\ \bibinfo {author} {\bibfnamefont {E.}~\bibnamefont
  {Mart\'{\i}n-Mart\'{\i}nez}},\ }\bibfield  {title} {\bibinfo {title}
  {Sabotaging the harvesting of correlations from quantum fields},\ }\href
  {https://doi.org/10.1103/PhysRevD.105.065011} {\bibfield  {journal} {\bibinfo
   {journal} {Phys. Rev. D}\ }\textbf {\bibinfo {volume} {105}},\ \bibinfo
  {pages} {065011} (\bibinfo {year} {2022})}\BibitemShut {NoStop}%
\bibitem [{\citenamefont {Bueley}\ \emph {et~al.}(2022)\citenamefont {Bueley},
  \citenamefont {Huang}, \citenamefont {Gallock-Yoshimura},\ and\ \citenamefont
  {Mann}}]{Bueley2022}%
  \BibitemOpen
  \bibfield  {author} {\bibinfo {author} {\bibfnamefont {K.}~\bibnamefont
  {Bueley}}, \bibinfo {author} {\bibfnamefont {L.}~\bibnamefont {Huang}},
  \bibinfo {author} {\bibfnamefont {K.}~\bibnamefont {Gallock-Yoshimura}},\
  and\ \bibinfo {author} {\bibfnamefont {R.~B.}\ \bibnamefont {Mann}},\
  }\bibfield  {title} {\bibinfo {title} {Harvesting mutual information from
  {BTZ} black hole spacetime},\ }\href
  {https://doi.org/10.1103/PhysRevD.106.025010} {\bibfield  {journal} {\bibinfo
   {journal} {Phys. Rev. D}\ }\textbf {\bibinfo {volume} {106}},\ \bibinfo
  {pages} {025010} (\bibinfo {year} {2022})}\BibitemShut {NoStop}%
\bibitem [{\citenamefont {Mendez-Avalos}\ \emph {et~al.}(2022)\citenamefont
  {Mendez-Avalos}, \citenamefont {Henderson}, \citenamefont
  {Gallock-Yoshimura},\ and\ \citenamefont {Mann}}]{MendezAvalos2022}%
  \BibitemOpen
  \bibfield  {author} {\bibinfo {author} {\bibfnamefont {D.}~\bibnamefont
  {Mendez-Avalos}}, \bibinfo {author} {\bibfnamefont {L.~J.}\ \bibnamefont
  {Henderson}}, \bibinfo {author} {\bibfnamefont {K.}~\bibnamefont
  {Gallock-Yoshimura}},\ and\ \bibinfo {author} {\bibfnamefont {R.~B.}\
  \bibnamefont {Mann}},\ }\bibfield  {title} {\bibinfo {title} {Entanglement
  harvesting of three {U}nruh-{D}e{W}itt detectors},\ }\href
  {https://doi.org/10.1007/s10714-022-02956-x} {\bibfield  {journal} {\bibinfo
  {journal} {Gen. Relativ. Gravit.}\ }\textbf {\bibinfo {volume} {54}},\
  \bibinfo {pages} {87} (\bibinfo {year} {2022})}\BibitemShut {NoStop}%
\bibitem [{\citenamefont {Unruh}(1976)}]{Unruh1976}%
  \BibitemOpen
  \bibfield  {author} {\bibinfo {author} {\bibfnamefont {W.~G.}\ \bibnamefont
  {Unruh}},\ }\bibfield  {title} {\bibinfo {title} {Notes on black-hole
  evaporation},\ }\href {https://doi.org/10.1103/PhysRevD.14.870} {\bibfield
  {journal} {\bibinfo  {journal} {Phys. Rev. D}\ }\textbf {\bibinfo {volume}
  {14}},\ \bibinfo {pages} {870} (\bibinfo {year} {1976})}\BibitemShut
  {NoStop}%
\bibitem [{\citenamefont {DeWitt}(1979)}]{DeWitt}%
  \BibitemOpen
  \bibfield  {author} {\bibinfo {author} {\bibfnamefont {B.}~\bibnamefont
  {DeWitt}},\ }in\ \href@noop {} {\emph {\bibinfo {booktitle} {General
  Relativity: an Einstein Centenary Survey}}},\ \bibinfo {editor} {edited by\
  \bibinfo {editor} {\bibnamefont {{{Hawking}, S. and {Israel}, W.}}}}\
  (\bibinfo  {publisher} {Cambridge University Press, Cambridge},\ \bibinfo
  {year} {1979})\BibitemShut {NoStop}%
\bibitem [{\citenamefont {Mart\'{\i}n-Mart\'{\i}nez}\ \emph
  {et~al.}(2013)\citenamefont {Mart\'{\i}n-Mart\'{\i}nez}, \citenamefont
  {Montero},\ and\ \citenamefont {del Rey}}]{eduardoOld}%
  \BibitemOpen
  \bibfield  {author} {\bibinfo {author} {\bibfnamefont {E.}~\bibnamefont
  {Mart\'{\i}n-Mart\'{\i}nez}}, \bibinfo {author} {\bibfnamefont
  {M.}~\bibnamefont {Montero}},\ and\ \bibinfo {author} {\bibfnamefont
  {M.}~\bibnamefont {del Rey}},\ }\bibfield  {title} {\bibinfo {title}
  {Wavepacket detection with the {U}nruh-{D}e{W}itt model},\ }\href
  {https://doi.org/10.1103/PhysRevD.87.064038} {\bibfield  {journal} {\bibinfo
  {journal} {Phys. Rev. D}\ }\textbf {\bibinfo {volume} {87}},\ \bibinfo
  {pages} {064038} (\bibinfo {year} {2013})}\BibitemShut {NoStop}%
\bibitem [{\citenamefont {Mart\'{i}n-Mart\'{i}nez}\ and\ \citenamefont
  {Rodriguez-Lopez}(2018)}]{eduardo}%
  \BibitemOpen
  \bibfield  {author} {\bibinfo {author} {\bibfnamefont {E.}~\bibnamefont
  {Mart\'{i}n-Mart\'{i}nez}}\ and\ \bibinfo {author} {\bibfnamefont
  {P.}~\bibnamefont {Rodriguez-Lopez}},\ }\bibfield  {title} {\bibinfo {title}
  {Relativistic quantum optics: The relativistic invariance of the light-matter
  interaction models},\ }\href {https://doi.org/10.1103/PhysRevD.97.105026}
  {\bibfield  {journal} {\bibinfo  {journal} {Phys. Rev. D}\ }\textbf {\bibinfo
  {volume} {97}},\ \bibinfo {pages} {105026} (\bibinfo {year}
  {2018})}\BibitemShut {NoStop}%
\bibitem [{\citenamefont {Lopp}\ and\ \citenamefont
  {Mart\'{i}n-Mart\'{i}nez}(2021)}]{richard}%
  \BibitemOpen
  \bibfield  {author} {\bibinfo {author} {\bibfnamefont {R.}~\bibnamefont
  {Lopp}}\ and\ \bibinfo {author} {\bibfnamefont {E.}~\bibnamefont
  {Mart\'{i}n-Mart\'{i}nez}},\ }\bibfield  {title} {\bibinfo {title} {Quantum
  delocalization, gauge, and quantum optics: Light-matter interaction in
  relativistic quantum information},\ }\href
  {https://doi.org/10.1103/PhysRevA.103.013703} {\bibfield  {journal} {\bibinfo
   {journal} {Phys. Rev. A}\ }\textbf {\bibinfo {volume} {103}},\ \bibinfo
  {pages} {013703} (\bibinfo {year} {2021})}\BibitemShut {NoStop}%
\bibitem [{\citenamefont {Sab\'{\i}n}\ \emph {et~al.}(2010)\citenamefont
  {Sab\'{\i}n}, \citenamefont {Garc\'{\i}a-Ripoll}, \citenamefont {Solano},\
  and\ \citenamefont {Le\'on}}]{Sabin2010}%
  \BibitemOpen
  \bibfield  {author} {\bibinfo {author} {\bibfnamefont {C.}~\bibnamefont
  {Sab\'{\i}n}}, \bibinfo {author} {\bibfnamefont {J.~J.}\ \bibnamefont
  {Garc\'{\i}a-Ripoll}}, \bibinfo {author} {\bibfnamefont {E.}~\bibnamefont
  {Solano}},\ and\ \bibinfo {author} {\bibfnamefont {J.}~\bibnamefont
  {Le\'on}},\ }\bibfield  {title} {\bibinfo {title} {Dynamics of entanglement
  via propagating microwave photons},\ }\href
  {https://doi.org/10.1103/PhysRevB.81.184501} {\bibfield  {journal} {\bibinfo
  {journal} {Phys. Rev. B}\ }\textbf {\bibinfo {volume} {81}},\ \bibinfo
  {pages} {184501} (\bibinfo {year} {2010})}\BibitemShut {NoStop}%
\bibitem [{\citenamefont {Sab\'{\i}n}\ \emph {et~al.}(2012)\citenamefont
  {Sab\'{\i}n}, \citenamefont {Peropadre}, \citenamefont {del Rey},\ and\
  \citenamefont {Mart\'{\i}n-Mart\'{\i}nez}}]{Sabin2012}%
  \BibitemOpen
  \bibfield  {author} {\bibinfo {author} {\bibfnamefont {C.}~\bibnamefont
  {Sab\'{\i}n}}, \bibinfo {author} {\bibfnamefont {B.}~\bibnamefont
  {Peropadre}}, \bibinfo {author} {\bibfnamefont {M.}~\bibnamefont {del Rey}},\
  and\ \bibinfo {author} {\bibfnamefont {E.}~\bibnamefont
  {Mart\'{\i}n-Mart\'{\i}nez}},\ }\bibfield  {title} {\bibinfo {title}
  {Extracting {P}ast-{F}uture {V}acuum {C}orrelations {U}sing {C}ircuit
  {QED}},\ }\href {https://doi.org/10.1103/PhysRevLett.109.033602} {\bibfield
  {journal} {\bibinfo  {journal} {Phys. Rev. Lett.}\ }\textbf {\bibinfo
  {volume} {109}},\ \bibinfo {pages} {033602} (\bibinfo {year}
  {2012})}\BibitemShut {NoStop}%
\bibitem [{\citenamefont {Borrelli}\ \emph {et~al.}(2012)\citenamefont
  {Borrelli}, \citenamefont {Sab{\'{\i}}n}, \citenamefont {Adesso},
  \citenamefont {Plastina},\ and\ \citenamefont {Maniscalco}}]{Borrelli2012}%
  \BibitemOpen
  \bibfield  {author} {\bibinfo {author} {\bibfnamefont {M.}~\bibnamefont
  {Borrelli}}, \bibinfo {author} {\bibfnamefont {C.}~\bibnamefont
  {Sab{\'{\i}}n}}, \bibinfo {author} {\bibfnamefont {G.}~\bibnamefont
  {Adesso}}, \bibinfo {author} {\bibfnamefont {F.}~\bibnamefont {Plastina}},\
  and\ \bibinfo {author} {\bibfnamefont {S.}~\bibnamefont {Maniscalco}},\
  }\bibfield  {title} {\bibinfo {title} {Dynamics of atom{\textendash}atom
  correlations in the {F}ermi problem},\ }\href
  {https://doi.org/10.1088/1367-2630/14/10/103010} {\bibfield  {journal}
  {\bibinfo  {journal} {New J. Phys.}\ }\textbf {\bibinfo {volume} {14}},\
  \bibinfo {pages} {103010} (\bibinfo {year} {2012})}\BibitemShut {NoStop}%
\bibitem [{\citenamefont {Forn-D\'{i}az}\ \emph {et~al.}(2017)\citenamefont
  {Forn-D\'{i}az}, \citenamefont {Garc\'{i}a-Ripoll}, \citenamefont
  {Peropadre}, \citenamefont {Orgiazzi}, \citenamefont {Yurtalan},
  \citenamefont {Belyansky}, \citenamefont {Wilson},\ and\ \citenamefont
  {Lupascu}}]{FornDiaz2017}%
  \BibitemOpen
  \bibfield  {author} {\bibinfo {author} {\bibfnamefont {P.}~\bibnamefont
  {Forn-D\'{i}az}}, \bibinfo {author} {\bibfnamefont {J.~J.}\ \bibnamefont
  {Garc\'{i}a-Ripoll}}, \bibinfo {author} {\bibfnamefont {B.}~\bibnamefont
  {Peropadre}}, \bibinfo {author} {\bibfnamefont {J.-L.}\ \bibnamefont
  {Orgiazzi}}, \bibinfo {author} {\bibfnamefont {M.~A.}\ \bibnamefont
  {Yurtalan}}, \bibinfo {author} {\bibfnamefont {R.}~\bibnamefont {Belyansky}},
  \bibinfo {author} {\bibfnamefont {C.~M.}\ \bibnamefont {Wilson}},\ and\
  \bibinfo {author} {\bibfnamefont {A.}~\bibnamefont {Lupascu}},\ }\bibfield
  {title} {\bibinfo {title} {Ultrastrong coupling of a single artificial atom
  to an electromagnetic continuum in the nonperturbative regime},\ }\href
  {https://doi.org/10.1038/nphys3905} {\bibfield  {journal} {\bibinfo
  {journal} {Nat. Phys.}\ }\textbf {\bibinfo {volume} {13}},\ \bibinfo {pages}
  {39} (\bibinfo {year} {2017})}\BibitemShut {NoStop}%
\bibitem [{\citenamefont {Ardenghi}(2018)}]{Ardenghi2018}%
  \BibitemOpen
  \bibfield  {author} {\bibinfo {author} {\bibfnamefont {J.~S.}\ \bibnamefont
  {Ardenghi}},\ }\bibfield  {title} {\bibinfo {title} {Entanglement harvesting
  in double-layer graphene by vacuum fluctuations in a microcavity},\ }\href
  {https://doi.org/10.1103/PhysRevD.98.045006} {\bibfield  {journal} {\bibinfo
  {journal} {Phys. Rev. D}\ }\textbf {\bibinfo {volume} {98}},\ \bibinfo
  {pages} {045006} (\bibinfo {year} {2018})}\BibitemShut {NoStop}%
\bibitem [{\citenamefont {Janzen}\ \emph {et~al.}(2022)\citenamefont {Janzen},
  \citenamefont {Dai}, \citenamefont {Ren}, \citenamefont {Shi},\ and\
  \citenamefont {Lupascu}}]{Janzen2022}%
  \BibitemOpen
  \bibfield  {author} {\bibinfo {author} {\bibfnamefont {N.}~\bibnamefont
  {Janzen}}, \bibinfo {author} {\bibfnamefont {X.}~\bibnamefont {Dai}},
  \bibinfo {author} {\bibfnamefont {S.}~\bibnamefont {Ren}}, \bibinfo {author}
  {\bibfnamefont {J.}~\bibnamefont {Shi}},\ and\ \bibinfo {author}
  {\bibfnamefont {A.}~\bibnamefont {Lupascu}},\ }\href@noop {} {\bibinfo
  {title} {Tunable coupler for mediating interactions between a two-level
  system and a waveguide from a decoupled state to the ultra-strong coupling
  regime}} (\bibinfo {year} {2022}),\ \Eprint
  {https://arxiv.org/abs/2208.05571} {arXiv:2208.05571 [quant-ph]} \BibitemShut
  {NoStop}%
\bibitem [{\citenamefont {Salton}\ \emph
  {et~al.}(2015{\natexlab{b}})\citenamefont {Salton}, \citenamefont {Mann},\
  and\ \citenamefont {Menicucci}}]{Salton2015}%
  \BibitemOpen
  \bibfield  {author} {\bibinfo {author} {\bibfnamefont {G.}~\bibnamefont
  {Salton}}, \bibinfo {author} {\bibfnamefont {R.~B.}\ \bibnamefont {Mann}},\
  and\ \bibinfo {author} {\bibfnamefont {N.~C.}\ \bibnamefont {Menicucci}},\
  }\bibfield  {title} {\bibinfo {title} {Acceleration-assisted entanglement
  harvesting and rangefinding},\ }\href
  {https://doi.org/10.1088/1367-2630/17/3/035001} {\bibfield  {journal}
  {\bibinfo  {journal} {New J. Phys.}\ }\textbf {\bibinfo {volume} {17}},\
  \bibinfo {pages} {035001} (\bibinfo {year} {2015}{\natexlab{b}})}\BibitemShut
  {NoStop}%
\bibitem [{\citenamefont {Zhang}\ and\ \citenamefont {Yu}(2020)}]{Zhang2020}%
  \BibitemOpen
  \bibfield  {author} {\bibinfo {author} {\bibfnamefont {J.}~\bibnamefont
  {Zhang}}\ and\ \bibinfo {author} {\bibfnamefont {H.}~\bibnamefont {Yu}},\
  }\bibfield  {title} {\bibinfo {title} {Entanglement harvesting for
  {U}nruh-{D}e{W}itt detectors in circular motion},\ }\href
  {https://doi.org/10.1103/PhysRevD.102.065013} {\bibfield  {journal} {\bibinfo
   {journal} {Phys. Rev. D}\ }\textbf {\bibinfo {volume} {102}},\ \bibinfo
  {pages} {065013} (\bibinfo {year} {2020})}\BibitemShut {NoStop}%
\bibitem [{\citenamefont {Fewster}\ and\ \citenamefont
  {Verch}(2020)}]{fewster1}%
  \BibitemOpen
  \bibfield  {author} {\bibinfo {author} {\bibfnamefont {C.~J.}\ \bibnamefont
  {Fewster}}\ and\ \bibinfo {author} {\bibfnamefont {R.}~\bibnamefont
  {Verch}},\ }\bibfield  {title} {\bibinfo {title} {Quantum {F}ields and
  {L}ocal {M}easurements},\ }\href {https://doi.org/10.1007/s00220-020-03800-6}
  {\bibfield  {journal} {\bibinfo  {journal} {Commun. Math. Phys.}\ }\textbf
  {\bibinfo {volume} {378}},\ \bibinfo {pages} {851} (\bibinfo {year}
  {2020})}\BibitemShut {NoStop}%
\bibitem [{\citenamefont {Fewster}(2019)}]{fewster2}%
  \BibitemOpen
  \bibfield  {author} {\bibinfo {author} {\bibfnamefont {C.~J.}\ \bibnamefont
  {Fewster}},\ }\href@noop {} {\bibinfo {title} {A generally covariant
  measurement scheme for quantum field theory in curved spacetimes}} (\bibinfo
  {year} {2019}),\ \Eprint {https://arxiv.org/abs/1904.06944} {arXiv:1904.06944
  [gr-qc]} \BibitemShut {NoStop}%
\bibitem [{\citenamefont {Bostelmann}\ \emph {et~al.}(2021)\citenamefont
  {Bostelmann}, \citenamefont {Fewster},\ and\ \citenamefont
  {Ruep}}]{fewster3}%
  \BibitemOpen
  \bibfield  {author} {\bibinfo {author} {\bibfnamefont {H.}~\bibnamefont
  {Bostelmann}}, \bibinfo {author} {\bibfnamefont {C.~J.}\ \bibnamefont
  {Fewster}},\ and\ \bibinfo {author} {\bibfnamefont {M.~H.}\ \bibnamefont
  {Ruep}},\ }\bibfield  {title} {\bibinfo {title} {Impossible measurements
  require impossible apparatus},\ }\href
  {https://doi.org/10.1103/PhysRevD.103.025017} {\bibfield  {journal} {\bibinfo
   {journal} {Phys. Rev. D}\ }\textbf {\bibinfo {volume} {103}},\ \bibinfo
  {pages} {025017} (\bibinfo {year} {2021})}\BibitemShut {NoStop}%
\bibitem [{\citenamefont {Polo-G\'omez}\ \emph {et~al.}(2022)\citenamefont
  {Polo-G\'omez}, \citenamefont {Garay},\ and\ \citenamefont
  {Mart\'{\i}n-Mart\'{\i}nez}}]{Jose}%
  \BibitemOpen
  \bibfield  {author} {\bibinfo {author} {\bibfnamefont {J.}~\bibnamefont
  {Polo-G\'omez}}, \bibinfo {author} {\bibfnamefont {L.~J.}\ \bibnamefont
  {Garay}},\ and\ \bibinfo {author} {\bibfnamefont {E.}~\bibnamefont
  {Mart\'{\i}n-Mart\'{\i}nez}},\ }\bibfield  {title} {\bibinfo {title} {A
  detector-based measurement theory for quantum field theory},\ }\href
  {https://doi.org/10.1103/PhysRevD.105.065003} {\bibfield  {journal} {\bibinfo
   {journal} {Phys. Rev. D}\ }\textbf {\bibinfo {volume} {105}},\ \bibinfo
  {pages} {065003} (\bibinfo {year} {2022})}\BibitemShut {NoStop}%
\bibitem [{\citenamefont {Mart\'{i}n-Mart\'{i}nez}\ \emph
  {et~al.}(2020)\citenamefont {Mart\'{i}n-Mart\'{i}nez}, \citenamefont
  {Perche},\ and\ \citenamefont {de~S.~L.~Torres}}]{us}%
  \BibitemOpen
  \bibfield  {author} {\bibinfo {author} {\bibfnamefont {E.}~\bibnamefont
  {Mart\'{i}n-Mart\'{i}nez}}, \bibinfo {author} {\bibfnamefont {T.~R.}\
  \bibnamefont {Perche}},\ and\ \bibinfo {author} {\bibfnamefont
  {B.}~\bibnamefont {de~S.~L.~Torres}},\ }\bibfield  {title} {\bibinfo {title}
  {General relativistic quantum optics: Finite-size particle detector models in
  curved spacetimes},\ }\href {https://doi.org/10.1103/PhysRevD.101.045017}
  {\bibfield  {journal} {\bibinfo  {journal} {Phys. Rev. D}\ }\textbf {\bibinfo
  {volume} {101}},\ \bibinfo {pages} {045017} (\bibinfo {year}
  {2020})}\BibitemShut {NoStop}%
\bibitem [{\citenamefont {Mart\'{\i}n-Mart\'{\i}nez}\ \emph
  {et~al.}(2021)\citenamefont {Mart\'{\i}n-Mart\'{\i}nez}, \citenamefont
  {Perche},\ and\ \citenamefont {Torres}}]{us2}%
  \BibitemOpen
  \bibfield  {author} {\bibinfo {author} {\bibfnamefont {E.}~\bibnamefont
  {Mart\'{\i}n-Mart\'{\i}nez}}, \bibinfo {author} {\bibfnamefont {T.~R.}\
  \bibnamefont {Perche}},\ and\ \bibinfo {author} {\bibfnamefont {B.~d. S.~L.}\
  \bibnamefont {Torres}},\ }\bibfield  {title} {\bibinfo {title} {Broken
  covariance of particle detector models in relativistic quantum information},\
  }\href {https://doi.org/10.1103/PhysRevD.103.025007} {\bibfield  {journal}
  {\bibinfo  {journal} {Phys. Rev. D}\ }\textbf {\bibinfo {volume} {103}},\
  \bibinfo {pages} {025007} (\bibinfo {year} {2021})}\BibitemShut {NoStop}%
\bibitem [{\citenamefont {Knuth}(1976)}]{Knuth1976}%
  \BibitemOpen
  \bibfield  {author} {\bibinfo {author} {\bibfnamefont {D.~E.}\ \bibnamefont
  {Knuth}},\ }\bibfield  {title} {\bibinfo {title} {Big omicron and big omega
  and big theta},\ }\href {https://doi.org/10.1145/1008328.1008329} {\bibfield
  {journal} {\bibinfo  {journal} {ACM SIGACT News}\ }\textbf {\bibinfo {volume}
  {8}},\ \bibinfo {pages} {18–24} (\bibinfo {year} {1976})}\BibitemShut
  {NoStop}%
\bibitem [{\citenamefont {Nielsen}\ and\ \citenamefont
  {Chuang}(2010)}]{Nielsen2010}%
  \BibitemOpen
  \bibfield  {author} {\bibinfo {author} {\bibfnamefont {M.~A.}\ \bibnamefont
  {Nielsen}}\ and\ \bibinfo {author} {\bibfnamefont {I.~L.}\ \bibnamefont
  {Chuang}},\ }\href {https://doi.org/10.1017/CBO9780511976667} {\emph
  {\bibinfo {title} {Quantum Computation and Quantum Information}}}\ (\bibinfo
  {publisher} {Cambridge University Press},\ \bibinfo {year}
  {2010})\BibitemShut {NoStop}%
\bibitem [{\citenamefont {Maeso-Garc\'{\i}a}\ \emph {et~al.}(2022)\citenamefont
  {Maeso-Garc\'{\i}a}, \citenamefont {Perche},\ and\ \citenamefont
  {Mart\'{\i}n-Mart\'{\i}nez}}]{HectorTales}%
  \BibitemOpen
  \bibfield  {author} {\bibinfo {author} {\bibfnamefont {H.}~\bibnamefont
  {Maeso-Garc\'{\i}a}}, \bibinfo {author} {\bibfnamefont {T.~R.}\ \bibnamefont
  {Perche}},\ and\ \bibinfo {author} {\bibfnamefont {E.}~\bibnamefont
  {Mart\'{\i}n-Mart\'{\i}nez}},\ }\bibfield  {title} {\bibinfo {title}
  {{E}ntanglement harvesting: {D}etector gap and field mass optimization},\
  }\href {https://doi.org/10.1103/PhysRevD.106.045014} {\bibfield  {journal}
  {\bibinfo  {journal} {Phys. Rev. D}\ }\textbf {\bibinfo {volume} {106}},\
  \bibinfo {pages} {045014} (\bibinfo {year} {2022})}\BibitemShut {NoStop}%
\bibitem [{\citenamefont {Ruep}(2021)}]{Ruep2021}%
  \BibitemOpen
  \bibfield  {author} {\bibinfo {author} {\bibfnamefont {M.~H.}\ \bibnamefont
  {Ruep}},\ }\bibfield  {title} {\bibinfo {title} {Weakly coupled local
  particle detectors cannot harvest entanglement},\ }\href
  {https://doi.org/10.1088/1361-6382/ac1b08} {\bibfield  {journal} {\bibinfo
  {journal} {Class. Quantum Gravity}\ }\textbf {\bibinfo {volume} {38}},\
  \bibinfo {pages} {195029} (\bibinfo {year} {2021})}\BibitemShut {NoStop}%
\bibitem [{\citenamefont {L\"{u}ders}(1951)}]{Luders1951}%
  \BibitemOpen
  \bibfield  {author} {\bibinfo {author} {\bibfnamefont {G.}~\bibnamefont
  {L\"{u}ders}},\ }\bibfield  {title} {\bibinfo {title} {\"{U}ber die
  {Z}ustands\"{a}derung durch den {M}e{\ss}proze{\ss}},\ }\href
  {https://doi.org/10.1002/andp.20065180904} {\bibfield  {journal} {\bibinfo
  {journal} {Ann. Phys. (Leipzig)}\ }\textbf {\bibinfo {volume} {8}},\ \bibinfo
  {pages} {322} (\bibinfo {year} {1951})}\BibitemShut {NoStop}%
\bibitem [{\citenamefont {Hellwig}\ and\ \citenamefont
  {Kraus}(1970)}]{Hellwig1970formal}%
  \BibitemOpen
  \bibfield  {author} {\bibinfo {author} {\bibfnamefont {K.~E.}\ \bibnamefont
  {Hellwig}}\ and\ \bibinfo {author} {\bibfnamefont {K.}~\bibnamefont
  {Kraus}},\ }\bibfield  {title} {\bibinfo {title} {Formal {D}escription of
  {M}easurements in {L}ocal {Q}uantum {F}ield {T}heory},\ }\href
  {https://doi.org/10.1103/PhysRevD.1.566} {\bibfield  {journal} {\bibinfo
  {journal} {Phys. Rev. D}\ }\textbf {\bibinfo {volume} {1}},\ \bibinfo {pages}
  {566} (\bibinfo {year} {1970})}\BibitemShut {NoStop}%
\end{thebibliography}%
    
\end{document}